\documentstyle[12pt,epsf]{article}
\advance\voffset by -1.8cm
\advance\hoffset by -1.9cm
\def\be{\begin{equation}}
\def\ee{\end{equation}}

\def\bbbz {{\sf Z\!\!Z}}

\def\bbbc{{\mathchoice {\setbox0=\hbox{$\displaystyle\rm C$}\hbox{\hbox
to0pt{\kern0.4\wd0\vrule height0.9\ht0\hss}\box0}}
{\setbox0=\hbox{$\textstyle\rm C$}\hbox{\hbox
to0pt{\kern0.4\wd0\vrule height0.9\ht0\hss}\box0}}
{\setbox0=\hbox{$\scriptstyle\rm C$}\hbox{\hbox
to0pt{\kern0.4\wd0\vrule height0.9\ht0\hss}\box0}}
{\setbox0=\hbox{$\scriptscriptstyle\rm C$}\hbox{\hbox
to0pt{\kern0.4\wd0\vrule height0.9\ht0\hss}\box0}}}}

\def\pmb#1{\setbox0=\hbox{#1}%
 \kern-.025em\copy0\kern-\wd0
 \kern.05em\copy0\kern-\wd0
 \kern-.025em\raise.0433em\box0 }
\def\sq{\hbox{\rlap{$\sqcap$}$\sqcup$}}
\def\qed{\ifmmode\sq\else{\unskip\nobreak\hfil
\penalty50\hskip1em\null\nobreak\hfil\sq
\parfillskip=0pt\finalhyphendemerits=0\endgraf}\fi}

\def\bbbone {{\mathchoice {{\rm 1\mskip-4mu l}} {{\rm 1\mskip-4mu l}}
{{\rm 1\mskip-4.5mu l}} {{\rm 1\mskip-5mu l}}}}
\def\square{\mathchoice\sqr56\sqr56\sqr{2.7}4\sqr{1.9}4}
\def\sqr#1#2{{\vcenter{\vbox{\hrule height.#2pt
                        \hbox{\vrule width.#2pt height#1pt \kern#1pt
                        \vrule width.#2pt}
                        \hrule height.#2pt}}}}

\begin{document}

\thispagestyle{empty}
\def\thefootnote{\fnsymbol{footnote}}
\begin{flushright}
  HUTP-97/A046 \\
  MIT-CTP-2670 \\
  hep-th/9709013
\end{flushright} \vskip 2.0cm

\begin{center}\LARGE
{\bf Exceptional groups from open strings}
\end{center} \vskip 1.0cm
\begin{center}\large
       Matthias R.\ Gaberdiel%
       \footnote{E-mail  address: {\tt gaberd@string.harvard.edu}}%
       \footnote{Address from 1st September 1997: DAMTP, University of
       Cambridge, Cambridge, CB3 9EW, U.K.}
       and 
       Barton Zwiebach%
       \footnote{E-mail  address: {\tt zwiebach@irene.mit.edu}}%
       \footnote{Address from 1st September 1997: Center for Theoretical
       Physics, MIT, Cambridge, MA 02139}
       \end{center}
\vskip0.5cm

\begin{center}
Lyman Laboratory of Physics\\
Harvard University\\
Cambridge, MA 02138
\end{center}
\vskip 4em
\begin{center}
August 1997
\end{center}
\vskip 1cm
\begin{abstract}

We consider type IIB theory compactified on a two-sphere in the presence of
mutually nonlocal 7-branes. The BPS states associated with the
gauge vectors of exceptional groups are seen to arise from open strings
connecting the 7-branes, and multi-pronged open strings capable of ending
on more than two 7-branes. These multi-pronged strings are built from open
string junctions that arise naturally when strings cross 7-branes. The
different string configurations can be multiplied as traditional open
strings, and are shown to generate the structure of exceptional groups.  

\end{abstract}

\vfill

\setcounter{footnote}{0}
\def\thefootnote{\arabic{footnote}}
\newpage

\renewcommand{\theequation}{\thesection.\arabic{equation}}

\section{Introduction}
\setcounter{equation}{0}

It has been the experience of physicists that exceptional gauge groups do
not arise in the perturbative open string setting.  The traditional
Chan-Paton construction allows open strings to carry only unitary,
unitary-symplectic, and orthogonal gauge groups \cite{cpc}.  This
conclusion was corroborated recently in a general analysis based on
classical open string field theory.  The analysis established that if open
strings carry a symmetry structure without spacetime interpretation,
unitarity and ghost number constraints require the classical theory to be
on-shell equivalent to a theory with a Chan-Paton gauge group \cite{GZ1}.
\vskip4pt

Generalized Chan-Paton constructions are possible. Open superstring
theories with D-branes and/or orientifold projections can give rise to
somewhat more intricate gauge groups and matter content, but again, no
exceptional gauge groups were encountered \cite{mgc}.  We recently
classified a class of generalized Chan-Paton constructions starting with
the assumption that the open string state space decomposes into sectors
that multiply according to a semigroup \cite{GZ2}.  We showed that open
string consistency requires the semigroup to be a Brandt semigroup, and the
known classification of those semigroups indicates that such open string
theories have the structure of theories with D-branes, carrying possibly
some additional structure.  It is not clear whether or not this structure
could be used to generate novel theories of open strings.
\medskip

There exists, however, a natural setup where an open-string interpretation
for exceptional gauge groups has been investigated
\cite{dasgupta,johansen}.\footnote{In a different vein, the heterotic
$E_8\times E_8$ string has been recently constructed as a soliton of
type I  \cite{BD}.} This setup is that of IIB superstrings
compactified on a two-sphere $S^2$ in a background of parallel 7-branes
which extend in the uncompactified directions and are points on the
two-sphere.  Such compactifications can be viewed as F-theory
compactifications on an elliptically fibered K3, where the base is $S^2$
and the fiber is a two-torus $T^2$ \cite{vafa}. The complex structure of
the $T^2$ varies as one moves on the base, and the points where this
complex structure becomes singular represent the positions of the 7-branes
in the IIB context. Twenty four such points are required in order to have a
smooth K3, and it becomes singular only when the points coalesce. When this
happens, non-contractible two-cycles shrink to zero size. The singularities
of elliptic K3's can be of $A_n$-type, $D_n$-type or $E_n$-type, and are
labeled in this way because the intersection numbers between collapsing
two-cycles generate the Cartan matrices associated to the Lie algebras
$A_n, D_n,$ and $E_n$.
\medskip

An F-theory background with an $A_{n-1}$ singularity corresponds to a
configuration of $n$ mutually local 7-branes in IIB string theory,
giving rise to a $su(n)$ gauge theory by means of conventional open
strings ending on the 7-branes.  The case of a $D_4$ singularity was
considered very explicitly by Sen \cite{senorientifold}, who showed that in
the perturbative regime this theory was equivalent to a IIB orientifold
with four coincident D7 branes, and by a duality transformation, equivalent
to a Type I open string theory.  In the non-perturbative regime the
orientifold is resolved into two 7-branes, which are nonlocal with respect
to each other and the other four 7-branes.  Following the analysis of
Sen, F-theory backgrounds involving $E_6, E_7$, and $E_8$ singularities
were presented by Dasgupta and Mukhi \cite{dasgupta}. The explicit brane
description of the exceptional singularities as a IIB background of
mutually nonlocal 7-branes was given by Johansen \cite{johansen}, who also
described candidates for the open strings corresponding to the gauge
vectors of the resulting exceptional groups.
\medskip

In the above description of the $E_{n+1}$ gauge algebra, an 
$su(n)\times su(2)\times u(1)$ gauge subalgebra is realized manifestly by
{\it conventional} open strings stretching between mutually local
7-branes. The other generators of the exceptional algebra are believed to 
arise as {\it essentially conventional} open strings stretching along
nontrivial paths between
sometimes mutually nonlocal branes; this requires that the open string 
crosses suitable branch-cuts that convert the string into an
$\mbox{SL}(2,\bbbz)$ transform that can end on the final 7-brane. 
\smallskip

There are shortcomings, however, that prevent this from being a clear open
string description of exceptional groups. In the standard description, the
charges carried by an open string are determined by the branes it ends
on. In the above description the candidate open strings for the
non-manifest generators must carry charges of branes they do not end on.
Moreover, the multiplication of open strings does not work in a natural
way. Open strings corresponding to generators whose Lie bracket does not
vanish are seen not to have common endpoints that would allow one to
combine them in the usual fashion.
\smallskip

The main purpose of the present paper is to provide an open-string
description of exceptional groups that avoids these difficulties. We do not
modify the 7-brane configurations described above, but we propose that the
fundamental objects that are necessary include not only open strings having
two endpoints, but also multi-pronged open strings having more than two
endpoints. These $n$-pronged open strings are built from open string
junctions. Since they have $n$ free endpoints, they can 
be charged with respect to $n$ gauge groups. We will describe the
multi-pronged open strings representing manifestly the hitherto non-manifest
additional generators of the exceptional algebra, show that they have the
desired charges, and that they can be combined by joining the prongs in the
usual open string theory way. 
\smallskip

Actually, the multi-pronged open strings are naturally related to
conventional (two-pronged) open strings. A three-pronged string can arise
when an ordinary string looping around a 7-brane it cannot end on,
crosses the 7-brane, and in the process a one-brane or an extra prong is
created. This new prong is necessary for charge conservation and arises in
a way that is completely analogous to the way new branes are created 
in the Hanany-Witten effect \cite{hananywitten}. (In fact, the two
processes are related by U-duality, see also \cite{08}.) 
The two desciptions that can be obtained from one another by a
crossing of branes describe the same BPS state in different regions of
the moduli space of the positions of the 7-branes.
Three string junctions have been considered before by Schwarz
\cite{schwarz} who suggested that they represent BPS configurations and 
anticipated their physical relevance.
\smallskip

It is perhaps not too speculative to suggest that our results point to a
possible non-perturbative formulation of open string theory based on open
strings and their multi-pronged versions. The world sheet of a multi-pronged
string is a two dimensional manifold except at the world line of the common
endpoint.\footnote{Years ago J. Goldstone asked one of us why 
such worldsheets were not included in string theory.}  Moreover, in the
covariant open string theory of Witten \cite{Witten} where the string
midpoint is singled out, an open string is naturally a two-pronged string.
Finally, just as open string endpoints can join to form closed strings,
joining all endpoints of several $n_i$-pronged open strings give objects
that would look as polyhedral closed string junctions, objects that are
formally reminiscent of the closed string polyhedra defining classical
closed string field theory \cite{polyhedra}.
\medskip

The paper is organized as follows. In section~2 we introduce notation,
discuss open string junctions, and explain how they can arise as branes
cross. We also review the configurations of 7-branes that are necessary for
exceptional groups, and the embeddings of the perturbative subalgebras.
In section~3, we study some of the open string geodesics that represent BPS
states. The generators of the exceptional groups are constructed in
section~4, where we also analyze their multiplication, and  explain how
they are related to ordinary geodesics. Section~5 contains some conclusions
and open questions.

\section{Strings, 1-branes and 7-branes}
\setcounter{equation}{0}

\subsection{7-branes and monodromies}

Let us consider IIB string theory compactified on a two-sphere in the
presence of a set of parallel 7-branes which appear as points on the
two-sphere. The IIB theory has different 7-branes which are labeled by
$[p,q]$, where $p$ and $q$ are relatively prime. The theory also possesses
different strings which are similarly labeled by $({p\atop q})$, where
again $p$ and $q$ are relatively prime, and a $({p\atop q})$-string can end
on a $[p,q]$ 7-brane.\footnote{Since $({p\atop q})$ and $({-p\atop -q})$
strings only differ by orientation, the $({-p\atop -q})$ string can also
end on a $[p,q]$ 7-brane.}

We choose the convention that the elementary string is $({1\atop 0})$,  
and the D-string is $({0\atop1})$. The $({p\atop q})$ strings can then
be thought of as bound states of $p$ elementary strings and $q$ D-strings
\cite{wittenbound}. The ordinary D7-brane has labels $[1,0]$.

Suppose that an elementary string ends on an ordinary D7-brane. Using 
the $\mbox{SL}(2,\bbbz)$ transformation with the matrix $g_{p,q}$, we can
translate the string into a $({p\atop q})$ string 
\be
\pmatrix{p\cr q } =  g_{p,q} \cdot \pmatrix{1\cr 0} \,, \quad g_{p,q} = 
\pmatrix{p& r\cr q &s} , \,\, ps-qr=1\,. 
\ee
After this transformation the $({p\atop q})$ string ends on a 7-brane
which must be, by definition, a $[p,q]$ 7-brane.  We must therefore view a
$[p,q]$ 7-brane as the $\mbox{SL}(2,\bbbz)$ transform with $g_{p,q}$ of the
ordinary $[1,0]$ D7-brane.  Concretely, the transformation can be thought
of as the transformation of the background fields that define the brane.

As discussed in \cite{douglasli} the $\mbox{SL}(2,\bbbz)$ matrix $g_{p,q}$
is not uniquely determined by the integers $p,q$, but presumably this
non-uniqueness has no physical consequences.  This can be seen explicitly
as far as the monodromy associated to the $[p,q]$ 7-brane is concerned.
Indeed, going around an ordinary D7 brane induces an $\mbox{SL}(2,\bbbz)$
transformation of the doublet of background NS and RR antisymmetric tensors
via the matrix  
\be 
T= \pmatrix{1& 1\cr 0& 1} = M_{1,0}  
\ee 
which is the monodromy $M_{1,0}$ of the $[1,0]$ D7-brane. This then implies
that the monodromy matrix of the $[p,q]$ 7-brane is 
\be
\label{genmon}
M_{p,q} = g_{p,q}\,  T\,  g_{p,q}^{-1} = \pmatrix{1-pq& p^2\cr -q^2& 1+pq}\,,
\ee
which depends only on $p$ and $q$. It should also be noted that
$M_{p,q}=M_{-p,-q}$.

If we introduce the complex combination $\tau=a+ie^{-\phi}$ of the dilaton
field $\phi$ and the axion field $a$, then $\tau$ transforms under an
$\mbox{SL}(2,\bbbz)$ transformation $g$ as
\be
\tau \mapsto {a \tau + b \over c \tau + d} \,, \qquad \mbox{where} \qquad
g=\pmatrix{a & b\cr c & d} \,.
\ee
In order to keep track of the different monodromies we shall only draw the
two-sphere on which the 7-brane is a point, and we shall introduce
appropriate branch cuts. We shall choose the convention that if an
$({r\atop s})$ string crosses in an anti-clockwise direction the branch cut
of a $[p,q]$ 7-brane, it is turned into a $M_{p,q}\cdot ({r\atop s})$
string; this is described in figure~1.

\begin{figure}[htb]
\epsfysize=5cm
\centerline{\epsffile{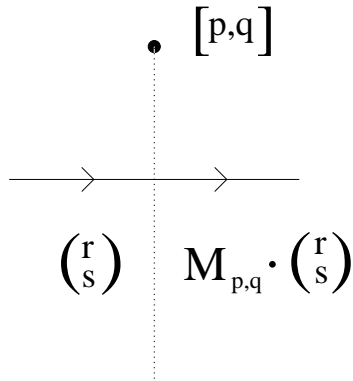}}
\caption{An $({r\atop s})$ string is shown crossing in the anti-clockwise
direction the branch cut associated to a $[p,q]$ 7-brane. The outgoing
string is the $M_{p,q} \cdot ({r\atop s})$ string.}
\end{figure}

It is simple to show that $({p\atop q})$ is the only eigenvector of
$M_{p,q}$, and that the corresponding eigenvalue is one. This is sensible
for it says that only a  $({p\atop q})$ string can go around a $[p,q]$
7-brane without being changed.

\subsection{Open string junctions}

Let us next analyze under which conditions three string junctions are
allowed. Suppose that three oriented open strings $({p_i\atop q_i})$,
$i=1,2,3$ form a three string junction, where, as always, $p_i$ and $q_i$
are relatively prime. As explained in \cite{schwarzreview,aharony}, charge
conservation implies then that the charges of the three strings must
satisfy 
\be
\label{chcons}
\begin{array}{lcl}
p_1+ p_2 + p_3  & =& 0\,, \\ 
q_1+ q_2 + q_3  & =& 0\,.
\end{array} 
\ee
It is however not clear whether this condition is already sufficient. In
fact, one may also want to require that a junction is only allowed if one of
the strings participating in the junction can end on another one, and in
the following we shall also impose this condition. It is known that 
the fundamental string $({\pm 1\atop 0})$ can end on the D-string
$({0\atop 1})$. Using the $\mbox{SL}(2,\bbbz)$ transformation 
$\pmatrix{p & r\cr q& s}$, this then implies that 
a $({\pm p\atop \pm q})$ string can end on a $({r\atop s})$ string.  We
therefore conclude  

\begin{verse}
{\it A} $({p\atop q})$ {\it string can end on a} $({r\atop s})$ 
{\it string if } $ps-qr = \pm 1$.
\end{verse}

If $({\pm p\atop \pm q})$ can end on $({r\atop s})$, then it can also end on
the outgoing $({p+r\atop q+s})$ string of the three string
junction (whose charges are determined by (\ref{chcons})). Furthermore, the
relation is symmetric: if $({p\atop q})$ can end on $({r\atop s})$, then
$({r\atop s})$ can end on $({p\atop q})$.  We shall therefore say that two
types of strings are {\it compatible} if they can end on one another.

It was suggested in \cite{schwarzreview} that under suitable conditions
on the geometry of the junction, the resulting configuration should be
BPS. 
\smallskip

Higher string junctions should also exist; for example, four-string
junctions should arise when the intermediate string joining two allowed
three-string junctions collapses.  Similar considerations should also apply
to higher string junctions. At present, we do not know if these are all the
allowed $n\geq 4$ string junctions.

\subsection{Brane crossings and creation of string junctions}

Next we want to explain that the three-string junction arises naturally
when a suitable string crosses a 7-brane. The effect is actually U-dual to
the Hanany-Witten effect \cite{hananywitten}; for the case of an D-string 
$({0 \atop 1})$ and an D7-brane $[1,0]$ this follows from the arguments in 
\cite{08}. The general case can then be deduced by $\mbox{SL}(2,Z)$
transformations. 

Suppose then that an $({r \atop s})$ string is crossing the branch cut of a 
$[p,q]$ 7-brane on which it is not allowed to end, {\it i.e.} 
$(r,s) \not= \pm (p,q)$. If the $({r \atop s})$ string is compatible with
the $({p \atop q})$ strings that can end on the 7-brane, {\it i.e.} if 
$ps-qr = e$, where $e =+1, \hbox{or}~, -1$, an interesting
possibility arises: as the $({r \atop s})$-string crosses the 7-brane, a
$({p \atop q})$-string is created that joins the $({r \atop s})$ string to
the 7-brane; this is represented pictorially in figure~2.

\begin{figure}[htb]
\epsfysize=5cm
\centerline{\epsffile{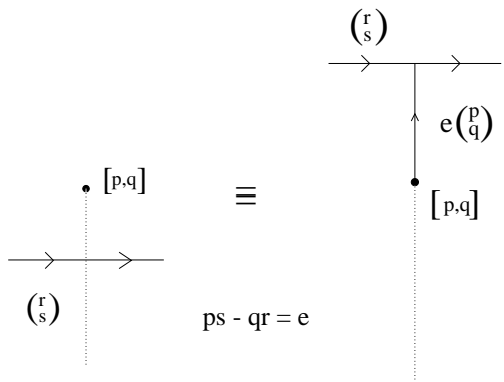}}
\caption{A 1-brane is created as the string crosses the 7-brane.}
\end{figure}

We should mention that the incoming and outgoing charges of the string are
the same for both representations. This follows from the equation
\be
M_{p,q} \cdot  \pmatrix{r\cr s} =  
\pmatrix{r\cr s}  + e  \pmatrix{p\cr q} \,,
\ee
which in turn can be derived with the help of (\ref{genmon}).
\smallskip

The two different representations are different descriptions of the same
BPS state; as in the Hanany-Witten effect they are valid in different
regions of the moduli space of positions of the 7-branes.

\subsection{Mutually nonlocal 7-branes}

In both the type IIB picture of compactification on $S^2$ and on the
equivalent F-theory picture of an elliptically fibered K3 with base $S^2$,
topological consistency conditions require a total of 24 7-branes.  As the
branes approach each other we get singularities that give rise to enhanced
gauge symmetries. 
\smallskip

In one familiar configuration the branes fall into four bunches of six
branes each, each bunch defining a $D_4$ singularity of K3 
and giving rise to an $so(8)$ gauge algebra \cite{senorientifold}.  
The six 7-branes on each bunch can be viewed as four ordinary $[1,0]$
D7-branes, to be denoted as $A$-branes, one $[3,-1]$ 7-brane ($B$-brane),
and one  $[1,-1]$ 7-brane ($C$-brane)\footnote{This assignment of $[p,q]$
labels to the  7-branes differs by an irrelevant overall SL(2,Z)
transformation from the assignments used in 
Refs.\cite{senorientifold,johansen}.}    
\be
\label{brdef}
A: [1,0]\,, \quad B: [3,-1]\,, \quad C: [1,-1] \,.
\ee
The corresponding monodromies, which we denote by abuse of notation as
$A$, $B$ and $C$, are obtained from (\ref{genmon}), and are given as
\be
A= \pmatrix{1&1\cr  0 & 1}\,, \quad B= \pmatrix{4&9\cr  -1 & -2}\,, \quad 
C= \pmatrix{2&1\cr  -1 & 0}\,.
\ee
Let us denote by $n_A, n_B,$ and $n_C$ the number of $A$, $B$ and $C$ branes,
respectively. Then we have for $so(8)$
\be
\label{soeight}
so(8) : \, n_A = 4, \quad n_B= 1\,, \quad n_C = 1\, .
\ee
As we encircle successively the $C$, $B$ and four $A$ branes,
we obtain an effective monodromy
\be
\label{basmon}
A^4 BC = -\bbbone\,.
\ee
\smallskip

All configurations of branes that will be relevant for us will have some
number of $A$, $B$ and $C$ branes. (In fact we shall always have $n_B=1$.)
The open strings joining the same types of branes along trivial paths
generate the gauge algebra $su(n_A)\times su(n_C)\times u(1)$. In the
following we shall mostly ignore the $u(1)$ factor which arises by some
mixing of the $u(1)$ factors that are associated to the different groups of
branes. 
\smallskip

For the different singularities on K3, the corresponding configurations of
(mutually non-local) 7-branes can be determined. As analyzed in
\cite{dasgupta} the branes can for example bunch into three groups of eight
branes, where each bunch consist of five $A$-branes, one $B$-brane
and two $C$-branes, and gives rise to the $E_6$ gauge 
algebra \cite{johansen}
\be
E_6 : \, n_A = 5, \quad n_B= 1\,, \quad n_C = 2\, .
\ee 
As we encircle successively the two $C$ branes, the $B$ brane,
and the five $A$ branes, we obtain an effective monodromy
\be
\label{e6mon}
A^5 BC^2 = \pmatrix{-1&-1\cr 1&0} = (ST)^2, \quad\hbox{with}
\quad S= \pmatrix{0&-1\cr1&0}\,. 
\ee
The monodromy $(ST)^2$ cubes to the identity, $[(ST)^2]^3=1$, and 
leaves $\tau = \exp(i\pi/3)$ invariant; we can thus chose
$\tau = \exp(i\pi/3)$  as a constant coupling.
\smallskip

The 24 branes can also fall into one bunch of six, and two bunches of nine
branes, giving rise to an $so(8)\times E_7\times E_7$ gauge algebra. One
can in this way realize $E_7$ with nine non-local branes \cite{johansen},
\be
E_7 :   \, n_A = 6, \quad n_B= 1\,, \quad n_C = 2\,,
\ee 
and with the associated monodromy
\be
A^6 BC^2 = S\,. 
\ee
This monodromy requires a (constant) string coupling $\tau=i$ for the
background. 
\smallskip 

Finally, one can group the 24 branes into a bunch of six, a bunch of eight,
and a bunch of ten giving rise to an $so(8)\times E_6\times E_8$ algebra.  
One can thus realize $E_8$ with ten non-local branes \cite{johansen}, 
\be
E_8 :  \, n_A = 7, \quad n_B= 1\,, \quad n_C = 2\, . 
\ee 
giving rise to an overall monodromy
\be
A^7 BC^2 = TS\,. 
\ee
The constant string coupling for this background is $\tau = \exp(i\pi/3)$.

\subsection{Exceptional Lie-algebras and  perturbative subalgebras}

In this section we will consider the Lie algebras $so(8), E_6, E_7$ and
$E_8$, and we will show, for each of them, how the adjoint representation 
transforms under the Lie subalgebra that can be realized manifestly with
the branes indicated in the previous section. The material in this section
is essentially an elaboration and explanation of some of the results of
Ref.\cite{johansen}.
\medskip

We begin with the case of  $so(8)$.  Here the manifest subalgebra
is $u(4)$ and therefore we consider $so(8) \to u(4) = su(4)\times u(1)$ where
the vector of $so(8)$ decomposes as $8 \to 4_{-1} + \overline 4_{+1}$.
The adjoint $28= (8\times 8)_a$ breaks as
\be
\label{so8br}
28 \to 15_0 + 1_0 + 6_{-2} + 6_{+2}\,.
\ee
The Lie algebra $so(8)$ can thus be viewed as generated by the elements
that generate $su(4)\times u(1)$ plus twelve other generators transforming
as $6_{-2} + 6_{+2}$.  The Lie bracket of $6_{-2}$ with  $6_{+2}$ gives
both the $15_0$ and the $1_0$.
\medskip

For $E_6$ we are interested in the subalgebra 
$su(5)\times su(2)\times u(1)$. This subalgebra is embedded in $E_6$ via
the maximal regular subalgebra $su(6)\times su(2)\subset E_6$.
For  $E_6 \to su(6)\times su(2)$
the adjoint decomposes as $78\to (35,1)+ (1,3)+ (20,2)$. Since for
$su(6) \to su(5)\times u(1)$, we have $6\to 5_1 + 1_{-5}$, we thus find
\begin{eqnarray}
\label{e6fdec}
78  &\to&   \,(24,1)_0 + (1,1)_0 + (1,3)_0 \nonumber \\ 
&{} &  + (10, 2)_{-3} + (\overline {10} , 2)_{3} \\
&{} &  + \,(\overline 5 , 1 )_{-6}\,\, + \,(5,1)_6  \,.\nonumber
\end{eqnarray} 
\medskip

For the case of $E_7$ we are interested in the manifest
subalgebra $su(6)\times su(2) \times u(1)$, which can be embedded into
$E_7$ via the maximal regular subalgebras $su(8)\subset E_7$, 
$so(12)\times su(2)\subset E_7$ or $su(6)\times su(3)\subset E_7$. 

The route via $su(8)$ uses the further decomposition 
$su(8)\to su(6)\times su(2)\times u(1)$ with $8\to (6,1)_1 + (1,2)_{-3}$.
In this decomposition, the adjoint of $E_7$ breaks into 
$su(6)\times su(2)\times u(1)$  representations which contain among others,
bifundamentals $(6,2)$ and $(\overline 6,2)$. Such representations cannot
be obtained with the mutually non-local branes we are considering, and
therefore this embedding of the desired subalgebra is not relevant for our
purposes. The embeddings of $su(6)\times su(2)\times u(1)$ in $E_7$
defined by using the last two maximal subalgebras give the same answer.
Under $E_7\to  so(12)\times su(2)$ we have $133\to (66,1)+(32',2)+(1,3)$,
and under $so(12)\to su(6)\times u(1)$ we have $12\to 6_1+ 6_{-1}$ 
and $32'\to 1_3+ 1_{-3}$ $ + 15_{-1} + \overline {15}_{+1}$, and we
therefore obtain
\begin{eqnarray}
\label{e7fdec}
133  &\to&   \,(35,1)_0 + (1,1)_0 + (1,3)_0 \nonumber \\ 
&{} &  + (15, 2)_{-1} + (\overline {15} , 2)_{1}  \nonumber \\
&{} &  + (\overline {15} , 1 )_{-2} + (15,1)_2   \\
&{} &  + \,(1 , 2)_{-3} \, + \, (1, 2)_{3} \,.  \nonumber
\end{eqnarray} 
\medskip

Finally let us consider the case of $E_8$, where the manifest
subalgebra is $su(7)\times su(2) \times u(1)$. This subalgebra can be
embedded into $E_8$ via the maximal regular subalgebras
$su(9)\subset E_8$ and $E_7\times su(2)\subset E_8$. In the first case, we
obtain again bi-fundamentals which do not 
arise in the brane configurations we are considering. In the second case,
the adjoint of $E_8$ decomposes as $248\to (133,1) + (1,3) + (56,2)$. 
We then have $E_7\to su(8)$ with $133\to 63+70$ and $56\to 28+\bar 28$,
and finally $su(8)\to su(7)\times u(1)$ with $8\to 7_1 + 1_{-7}$.  Using
this chain of embeddings we thus find that under 
$E_8 \to su(7)\times su(2) \times u(1)$ we have
\begin{eqnarray}
\label{e8fdec}
248  &\to&   \,(48,1)_0 + (1,1)_0 + (1,3)_0 \nonumber \\ 
&{} &  + (21,2)_2 + (\overline {21} , 2 )_{-2}  \nonumber \\
&{} &  +  (\overline{35}, 1)_{4} + (35 , 1)_{-4}     \\
&{} &  + (\overline 7,  2)_{6} + (\overline {7} , 2)_{-6}  \nonumber\\
&{} &  + (7,1)_8 + (\overline {7} , 1 )_{-8}\,. \nonumber 
\end{eqnarray} 
\medskip

This information can be conveniently organized in the following table. 
The first column contains the manifestly realized subalgebra, and the other
columns describe the representations of the additional generators. These
are either in the fundamental (2nd and 4th column) or the singlet (3rd and
5th column) of the $su(2)$ algebra. The representation of $su(n_A)$ is 
the completely antisymmetric 2nd, 4th, 6th rank tensor and the fundamental,
respectively. In the top line, the representations are labeled by 
boxes which are reminiscent of the Young tableaux of $su(n)\times su(2)$,
and we shall often use this notation below.

\vskip .1in
$$\hbox{\vbox{\offinterlineskip
\def\strut{\hbox{\vrule height 25pt depth 20pt width 0pt}}
\hrule
\halign{
\strut\vrule#\tabskip 0.1in&
\hfil$#$\hfil &
\vrule#&
\hfil$#$\hfil &
\vrule#&
\hfil$#$\hfil &
\vrule#&
\hfil$#$\hfil &
\vrule#&
\hfil$#$\hfil &
\vrule#&
\hfil$#$\hfil &
\vrule#\tabskip 0.0in\cr
&{} 
&&\hbox{Manifest subalgebra}
&& (\square^2_a, \square )
&& (\square^4_a, \cdot )
&& (\square^6_a, \square )
&& (\square, \cdot )
& \cr\noalign{\hrule}
&  E_6
&&su(5)\times su(2)\times u(1)
&& (10,2)_{-3}
&& (\overline 5, 1)_{-6} 
&& -- 
&& --
& \cr\noalign{\hrule}
&E_7
&&{su(6)\times su(2)\times u(1)}
&&(15,2)_{-1} 
&& (\overline{15},1)_{-2}
&& (1,2)_{-3}
&& --
& \cr\noalign{\hrule}
& E_8
&&{su(7)\times su(2)\times u(1)}
&&(21,2)_2
&& (\overline{35},1)_4 
&& (\overline 7,2)_6
&& (7,1)_8
&\cr\noalign{\hrule}
}}}$$

\section{Geodesics and BPS states}
\setcounter{equation}{0}

In the IIB superstring approach to exceptional gauge groups of
Ref.\cite{johansen}, the gauge vectors arise as geodesics on the
compactifying two-sphere whose endpoints are 7-branes. The
property of the different paths to be geodesics is somewhat difficult to
analyze, and it was only checked numerically in \cite{johansen}. 

In this section we shall take some small steps that improve
this situation. We will be able to show on general grounds that certain
classes of geodesics, for specific regions of the moduli of positions of
the 7-branes, exist. However, not even these (simple) classes of geodesics
seem to exist for completely arbitrary positions of the 7-branes. This is
in accord with our proposal that the conventional open strings do not
represent the relevant BPS states throughout the whole moduli spaces of
positions of the 7-branes. Rather, there exist various regions of the moduli
space where the relevant BPS states correspond to multi-pronged strings.

{}From the F-theory point of view, the ordinary open string geodesics
correspond to two-spheres with two marked points (which are the locations
of the 7-branes, where the torus degenerates). Our
$n$-pronged open string configurations are the natural generalization to
two-spheres with $n$ marked points; the locations of the open string
junctions themselves do not represent degenerate tori. 
Similar configurations have also been considered before in a slightly
different context in Ref.~\cite{OV}.

\subsection{Metrics on the two-sphere}

The geodesics that we aim to understand are geodesics on a two-sphere 
with a special metric. This metric was first obtained in \cite{greene},
and it is best described by using the F-theory picture of an elliptically
fibered K3 whose base is the $S^2$ in question. Taking $z$ to be the
complex coordinate on $S^2$, the K3 is described by the equation 
\be
\label{efibr}
y^2 = x^3 + f(z) x + g(z)\,,
\ee
which defines a torus with a complex structure for each value of $z$.  
Here $f$ and $g$ are polynomials of degree eight and twelve, respectively,
and the complex structure of the torus $\tau(z)$ is implicitly defined by the
equation 
\be
\label{modfix}
j(\tau(z)) = 4\cdot (24f)^3/\Delta\,,  \quad\hbox{with}\quad
 \Delta \equiv 27g^2+ 4f^3 = \prod_{i=1}^{24} (z-z_i)\,, 
\ee  
where the $z_i$'s are the positions of the 7-branes in the IIB description. 
Properly speaking, $\tau(z)$ is not a function, but defines a holomorphic
section in an $\mbox{SL}(2,\bbbz)$ bundle over the two-sphere. 

The metric on the $S^2$ is then given as
\be
ds^2 = \tau_2  \, \eta(\tau)^2 \bar\eta (\bar\tau)^2 \prod_i
(z-z_i)^{-1/12} (\bar z - \bar z_i)^{-1/12} \, dz d\bar z\,,
\ee
where $\tau_2=\mbox{Im}(\tau)$, and
\be
\label{expeta}
\eta^2 (\tau) = q^{1/12} \prod_{n=1}^\infty (1-q^n)^2\,, \quad q=\exp (2\pi
i\tau)\,, 
\ee
is the square of the Dedekind eta-function which satisfies
\be
\label{etatr}
\eta^2 (-1/\tau) = -i\tau \eta^2 (\tau) \,, \quad
\eta^2 (\tau+1) = e^{i\pi/6} \, \eta^2(\tau)\,.
\ee
It is straightforward to show that this metric is invariant under 
$\mbox{SL}(2,\bbbz)$ transformations of $\tau$. The masses of the states 
associated to $({p\atop q})$ strings must also take into account the string
tension $T_{p,q}$ of the $({p\atop q})$ string \cite{schwarz},
\be
T_{p,q} = {1\over \sqrt{\tau_2}}\, | p-q\tau \,|\,.
\ee
It is then natural to introduce the length element $ds_{p,q} = T_{p,q} ds$
which measures correctly the mass of the corresponding string, and the 
corresponding effective metric $ds^2_{p,q}$
\be
\label{effmet}
ds_{p,q}^2 = h(z)_{p,q} \,\overline{h_{p,q}(z)}\, dz d\bar z\, ,
\quad\hbox{with}\quad h_{p,q}(z) = (p-q\tau)\,
\eta^2(\tau) \prod_i (z-z_i)^{-1/12}\,.
\ee
Under an SL(2,Z) transformation we have
\be
\tau\to\tau'= {a\tau+ b\over c\tau + d}\, , 
\quad \pmatrix{p\cr q} \to \pmatrix{p'\cr q'}
=\pmatrix{a& b\cr
c& d}\cdot \pmatrix{p\cr q}\,,\,\,\,\hbox{and}\quad T_{p,q}\to T_{p',q'}\,,
\ee
and it follows that the effective metric is continuous across the branch
cuts emerging from the 7-branes, {\it i.e.} 
$ds_{p,q} (\tau) = ds_{p'q'}(\tau')$. Since the effective metric is the
modulus of an analytic one-form, it is flat, except for possible
singularities. 
\smallskip

The metric (\ref{effmet}) is typically rather complicated, but in some
cases the metric behaves (at least at large distance) as if it had a
conical singularity. By this one means that $h_{p,q}(z)$ is of the form 
\be
\label{asymp}
h_{p,q}(z) =  z^{-{\theta\over 2\pi}} (c + c' z + \cdots ) \,  ,  
\ee
where $\theta$ is the defect angle, and we have assumed, for simplicity,
that the conical singularity is at $z=0$. Here the expression
in parenthesis is simply the expansion of an analytic function that is
regular at the origin. In particular, the metric (\ref{effmet}) is of this
form if the term $(p-q\tau) \eta^2(\tau)$ is a constant (as a function of
$z$), and all 7-branes are located at $z_i=0$.
\smallskip

To proceed let us consider the configuration of a fundamental string in the
vicinity of a $[1,0]$ D7-brane, which, for convenience we place at
$z=0$. Since $M_{1,0}=T$, we have 
\be
\tau (z) \sim {1\over 2\pi i } \ln z\,.
\ee
and therefore $\eta^2 \sim z^{1\over 12}$. It then follows that the factor
$h_{1,0}(z) \sim \eta^2 z^{-{1\over 12}}$ is regular in this situation, and
the $({1\atop 0})$ string does not see a metric singularity at the position
of the $[1,0]$ D7-brane. This is, of course, as expected.

We can next consider the case where a $({1\atop 0})$ string loops around a 
collection of $N$ branes located in the vicinity of $z=0$.  The metric
(\ref{effmet}), far away from $z=0$ reads then
\be
h_{1,0}  \sim  \eta^2 (z)  z^{-{N\over 12}} \,.
\ee
If the collection of $N$ branes create an effective monodromy matrix
\be
\label{parmon}
M = \pmatrix{-1 & \, l \cr
0 & -1 } \,,
\ee
we have, as we go (anti-clockwise) around these branes, $\tau \to \tau-l$,
and therefore 
\be
\tau (z) \sim  {-l\over 2\pi i} \ln z\,, \quad \quad  \eta^2 (\tau(z)) \sim
z^{-{l\over 12}}\,. 
\ee
This implies that $h_{1,0} \sim z^{-{(N+l)\over 12}}$, and the group of 
$N$ branes creates a conical singularity that corresponds to a deficit
angle of $(N+l) 2\pi/12$.

\subsection{Indirect $AA$ and $CC$ geodesics}

The above considerations can be used to discuss the nature of $AA$ and $CC$
geodesics that encircle some number of 7-branes. As we shall see, in suitable
limits we can describe both kinds of geodesics as geodesics on a cone with
defect angle $\pi$. This is effectively an orientifold description.
\smallskip

Let us first consider the case of $AA$ geodesics. The different
(potential) geodesics that begin and end on an $A$ brane fall into two
classes, the {\it direct} and the {\it indirect} paths. The former are
all paths which do not cut any branch cuts, and the indirect paths are
those which begin on $A$, go around a $C$ brane, a $B$ brane, and then end
again on an $A$ brane. These geodesics are for example relevant for 
the brane configuration of $so(8)$ which requires four $A$ branes, one $B$
brane and one $C$ brane. In this case, if all six branes coincide in a
point, (\ref{basmon}) implies that the effective monodromy of the complete
configuration is minus the identity matrix. In the notation of
(\ref{parmon}) we then have $l=0$ and the metric for the fundamental string
has a conical singularity with defect angle $\pi$.   

As we remove the $A$ branes from the collapsed configuration the defect
angle for the $ds_{1,0}$ metric does not change. This follows for example
from the fact that an $A$ brane does not represent any singularity for the
fundamental $({1\atop 0})$ string. We can also check this explicitly, as 
$BC$ is of the form (\ref{parmon}) with $l=4$, and thus $N+l=2+4=6$ giving
again the defect angle of $\pi$. 
\smallskip

The situation for $CC$ strings is similar. Again there exist the direct
paths, and the indirect paths encircle four $A$ branes and one $B$
brane, whose effective monodromy is $A^4B = \pmatrix{0 &1\cr -1& -2}$. 
It is convenient to perform an $\mbox{SL}(2,\bbbz)$ transformation
$g= \pmatrix{1&0\cr -1& 1}$ which turns the $C$-brane into a conventional
$[1,0]$ D7-brane. The effective monodromy of the group of five 7-branes is
then $g^{-1} (A^4B) g = \pmatrix{-1&1\cr 0 & -1}$, which is of the form
(\ref{parmon}) with $l=1$. Altogether we therefore have again a conical
singularity with a defect angle of $\pi$ (as $N+l=5+1=6$). 
\medskip

The geometry of geodesics in a cone of defect angle $\pi$ is simple to
understand. Between any two branes located away from the apex there are
two geodesics, one `direct' geodesic which would become trivial should the
branes approach each other, and the `indirect' geodesic which goes around
the other side of the cone.\footnote{There are other geodesics that wind
several times around the apex, but they necessarily have self
intersections, and presumably do not give rise to independent BPS states.}
As long as the two $A$ branes are not contained on the same radial line
away from the apex, both geodesics avoid the apex.  On the other hand, the
indirect geodesics going from a brane to itself will necessarily go
through the apex, and their existence is therefore questionable. 
This suggests that for $n$ branes in the vicinity of a cone of defect
angle $\pi$, we get $n^2$ direct geodesics (distinguishing orientation) and
$n(n-1)$ indirect geodesics. 
\smallskip

This analysis applies directly only to the situation, where for the $AA$
strings, $B$ and $C$ coincide (and for the $CC$ string, the four $A$s and
$B$ coincide). If we separate $B$ and $C$, then the above picture is only
approximately true whenever we are far away from the region around the
apex. In the regions of moduli space where this approximation is not
valid, some of the geodesics are likely to become questionable, and it
seems plausible that open string junctions become the relevant objects.

\section{Exceptional groups}
\setcounter{equation}{0}

In this section we shall analyze the various different geodesics (and their
representatives involving string junctions) which are relevant for the
description of the exceptional groups. As mentioned earlier, the direct
geodesics between the $A$ branes and between the $C$ branes account for the
``manifest'' gauge subgroup. Here we shall only consider the additional
generators. We shall demonstrate that these generators have the
correct charges, and that they multiply correctly in order to account for
the structure of the exceptional groups.

\subsection{Versions of indirect $AA$ strings}

We shall start by considering the indirect $AA$ strings in more detail. As
explained before, these geodesics start on an $A$ brane, encircle a $C$
brane and a $B$ brane anti-clockwise, and end on a different $A$
brane; this is shown in Fig.~3 (a). From now on we shall always use the
convention that heavy dots represent $A$ branes, small circles represent
$B$ branes, and small squares represent $C$ branes. The dotted lines that
emanate from the 7-brane represent the corresponding branch cut, and the
strings that begin or end at a $[p,q]$ 7-brane are always $({p\atop q})$
strings.  

The configuration in Fig.~3 (a) is allowed as the monodromy of $C$ and $B$
transforms the $({1\atop 0})$ string that starts at the right $A$ brane
into a $BC \cdot ({1\atop 0})= ({-1\atop 0})$ string that arrives at the
left $A$ brane; this is equivalent to a $({1\atop 0})$ string departing
from the left $A$ brane. We note that both $A$ branes have departing
$({1\atop 0})$ strings.
\smallskip

As explained in the previous section, only those paths are geodesics, for
which the string begins and ends on different $A$ branes. This can
now also be understood pictorially: we can compose the strings in Fig.~3 (a)
with `direct' $AA$ strings (which represent the gauge bosons of $su(n_A)$),
and the fact that the string in Fig.~3 (a) departs from both $A$ branes
implies that the different strings transform as the antisymmetric tensor
representation of $su(n_A)$.

On the other hand, for fixed $A$ endpoints, there exist $n_C$ different
such configurations, depending on which $C$ brane is encircled. One 
should expect that the indirect $AA$ strings transform in the fundamental
representation of $su(n_C)$ (which is generated by the direct $CC$
strings), but since the configurations of Fig.~3 (a) do not have any
endpoints on a $C$ brane this is not manifest.

\begin{figure}[htb]
\vspace*{0.5cm}

\epsfysize=3.5cm
\epsffile{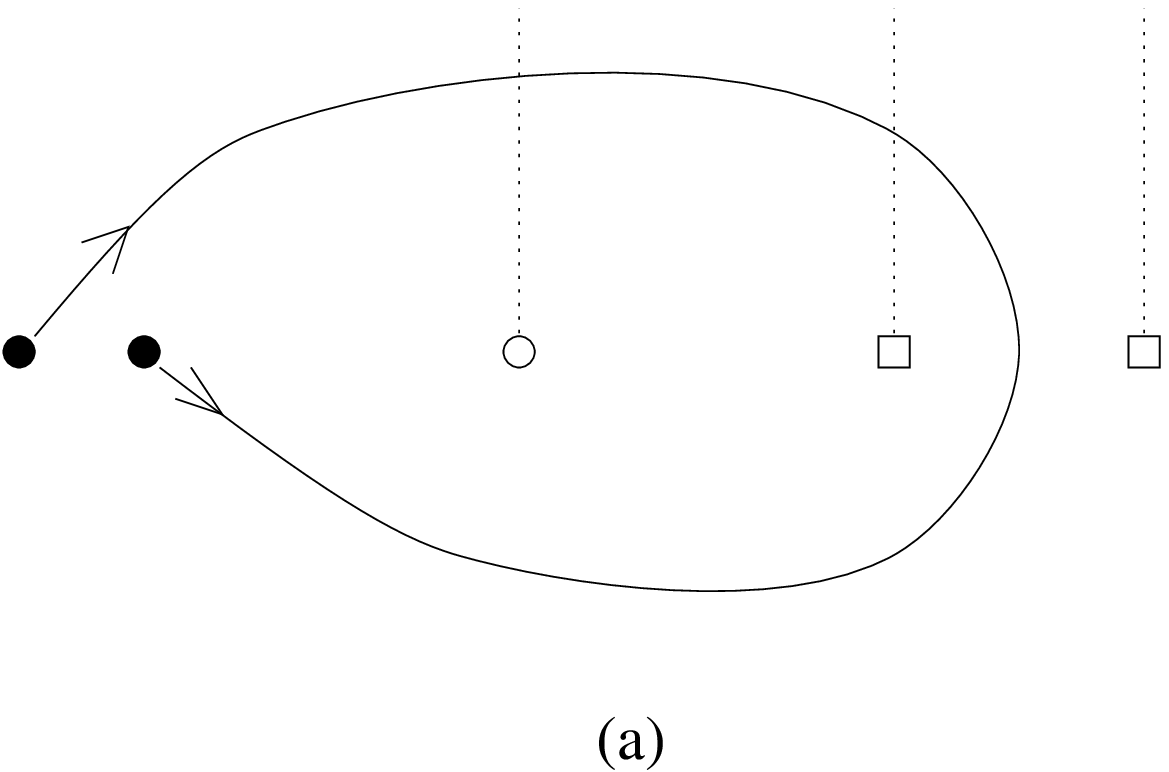}
\vspace*{-3.5cm}

\epsfysize=3.7cm
\hfill{\epsffile{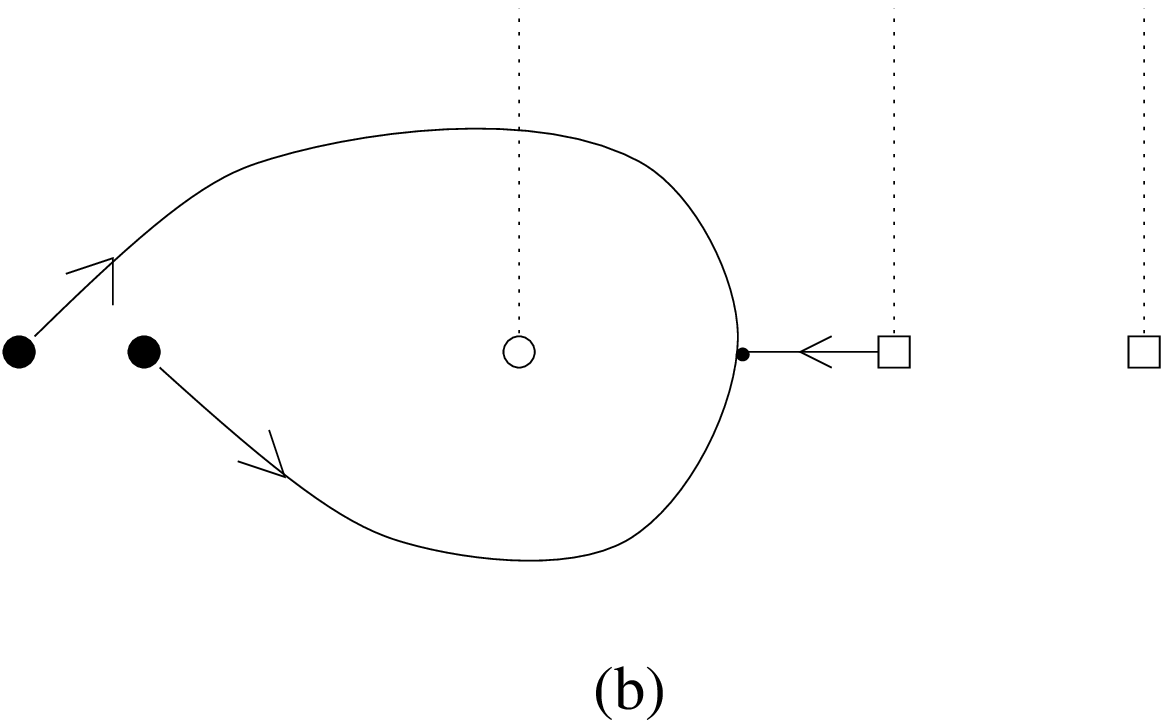}}
\vspace*{0.3cm}

\epsfysize=3.3cm
\epsffile{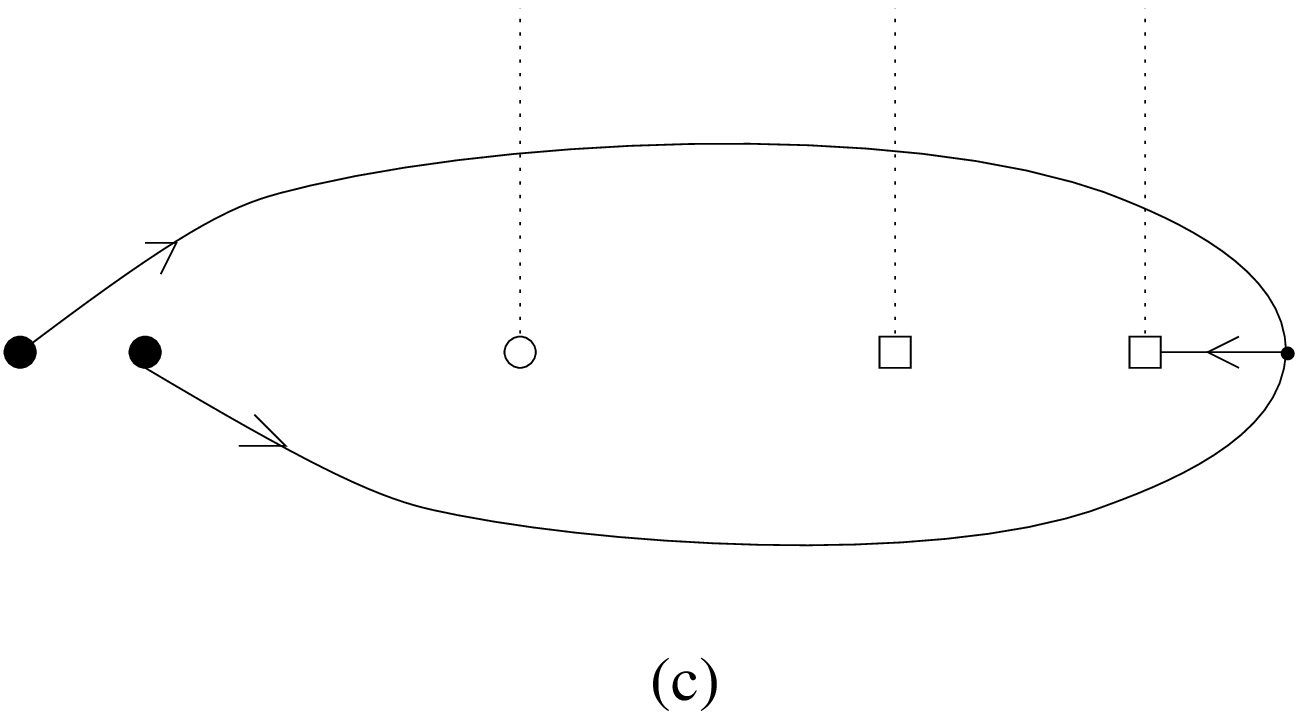}
\vspace*{-3.1cm}

\epsfysize=3.0cm
\hfill{\epsffile{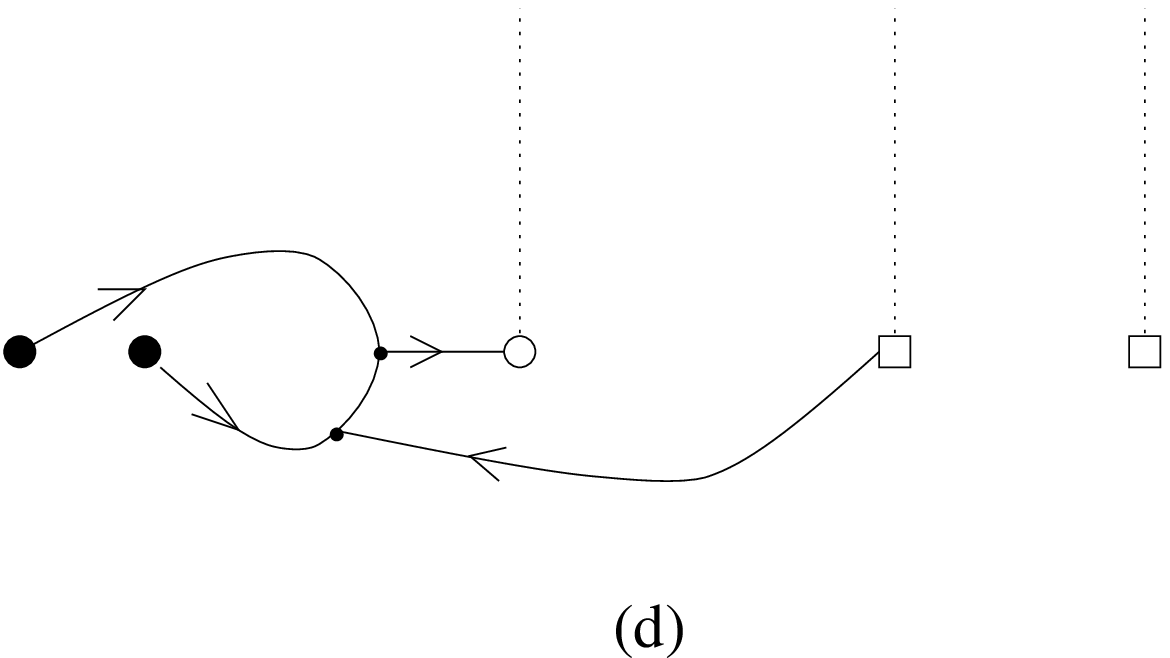}}

\epsfysize=3.7cm
\centerline{\epsffile{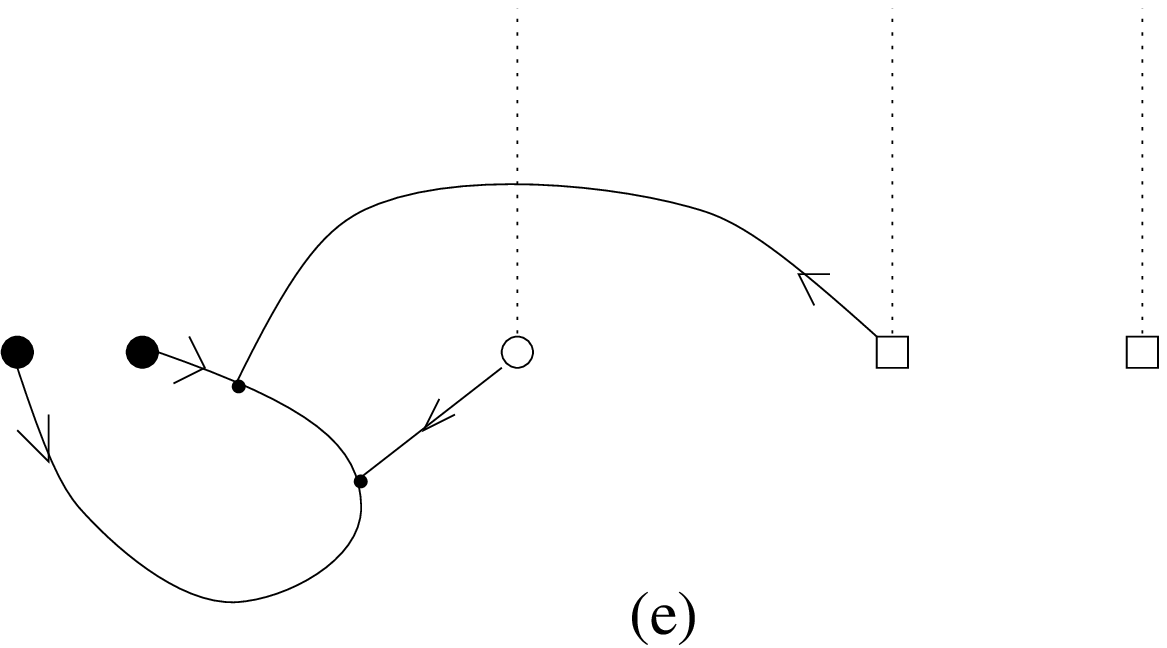}}
\caption{Five versions of the $AA$ geodesic which transforms as
$(\square^2_a, \square )$ of $su(n_A) \times su(n_C)$.}
\end{figure}

We can now, however, change the presentation by letting the open string cross
the $C$ brane it encloses. This is allowed by the rule described in Fig.~2
as the $({0\atop 1})$ string is compatible with the $({1\atop -1})$ string
associated to a $C$ brane. After crossing we get the extra open string
prong emerging from the $C$ brane, and we have obtained a three-string
junction as a representation of the BPS state; this is illustrated in
Fig.~3 (b). In this presentation the fact that the states transform as
fundamentals of $su(n_C)$ is now manifest, as we have an open string
ending on the $C$ brane, and the states compose naturally with direct $CC$
strings. 
\smallskip

Additional presentations are possible by doing further moves. We can go from
(a) to (c) by pushing the open string across the $C$ brane that is not
enclosed; this leads to a diagram where the prong goes into the $C$ brane. 
Two more presentations that are useful are shown in (d) and (e), both of
which follow by crossing the $B$ brane in presentation (b).  
\smallskip

The complex conjugate representation is described by exactly the same
diagrams with the exception that the arrows are reversed. Thus, for
example, while Fig.~3 (a) represents the $(10,2)_{-3}$ for the case of
$E_6$, the same figure with the $({1\atop 0})$ strings going into the $A$
branes would represent the $(\overline{10},2)_{3}$. This is
sensible on various accounts. First and foremost, the diagrams with 
arrows reversed are consistent if the original representations are
consistent: reversal of arrows is compatible with charge conservation at
junctions and with the crossing of branch cuts.  By construction they
represent the same number of states as the original representation, and
finally we can combine a representation with its complex conjugate by
gluing the string prongs at all 7-branes to obtain a singlet.

\subsection{Versions of the $CC$ strings}

As can be seen from the table given in section 2, the non-manifest
generators of $E_6$ include the $AA$ states considered above and
transforming as $(\square^2_a, \square)$ of $su(5)\times su(2)$, and, in
addition, states transforming as $(\square^4_a, \cdot)$. These are states
that can be represented as open strings beginning and ending on $C$ branes,
and enclosing four $A$ branes and one $B$ brane \cite{johansen}; this
geodesic is shown in Fig.~4 (a). A $({1\atop -1})$ string crosses a $B$ 
brane and four $A$ branes and in this process becomes a 
$A^4B \cdot ({1\atop -1}) = ({-1\atop +1})$ string ending on the other $C$
brane, or equivalently, an $({1\atop -1})$ departing from it. As explained
before in section~3, only those states are BPS which begin and end on two
different $C$ branes. This is also consistent with the fact that the states
in Fig.~4 (a) transform as the antisymmetric representation of
$su(n_C)$. As $n_C=2$ for the cases of $E_6$, $E_7$ and $E_8$, the indirect
$CC$ strings will be singlets of $su(2)$. 

\begin{figure}[htb]
\epsfysize=3.4cm
\epsffile{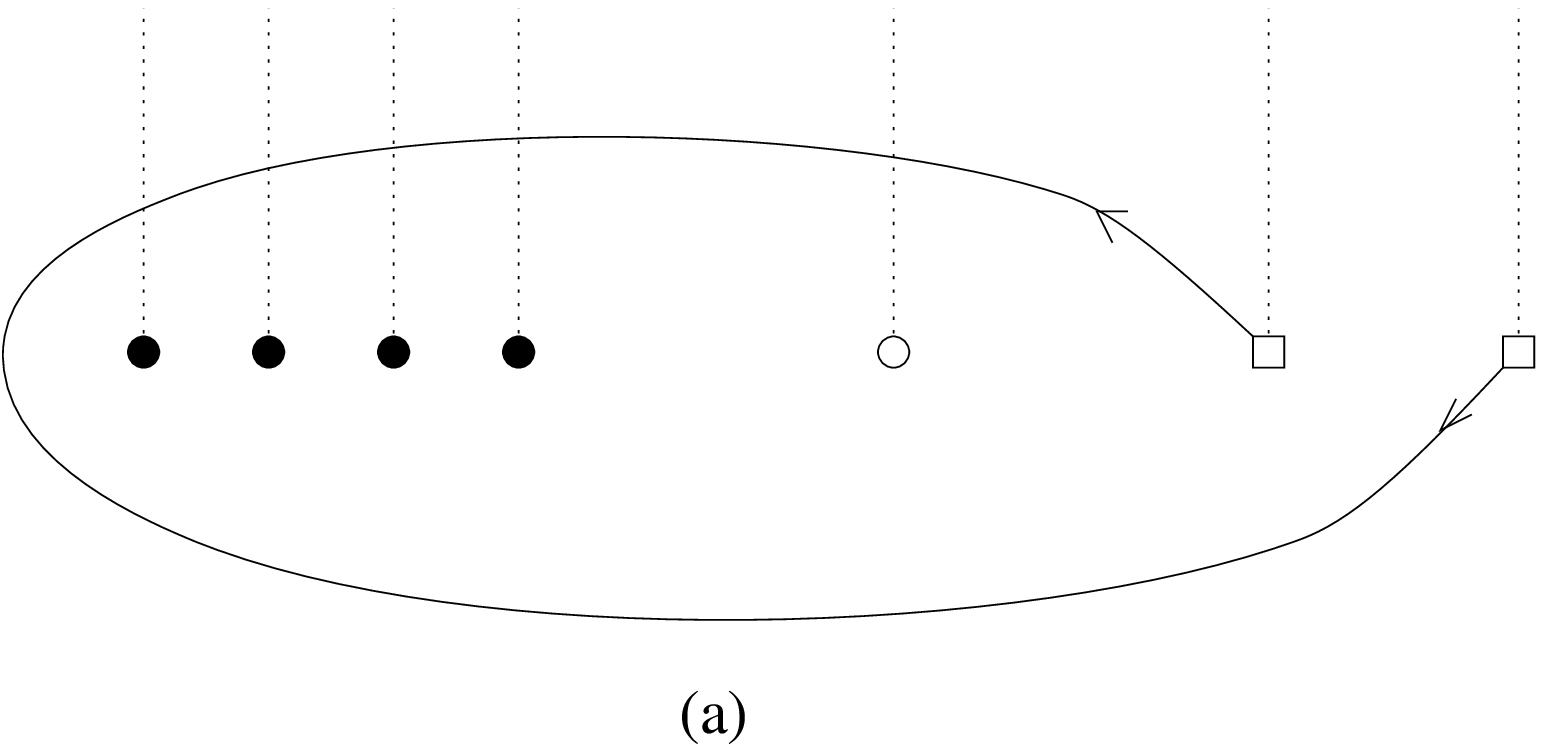}
\vspace*{-3.4cm}

\epsfysize=4.0cm
\hfill{\epsffile{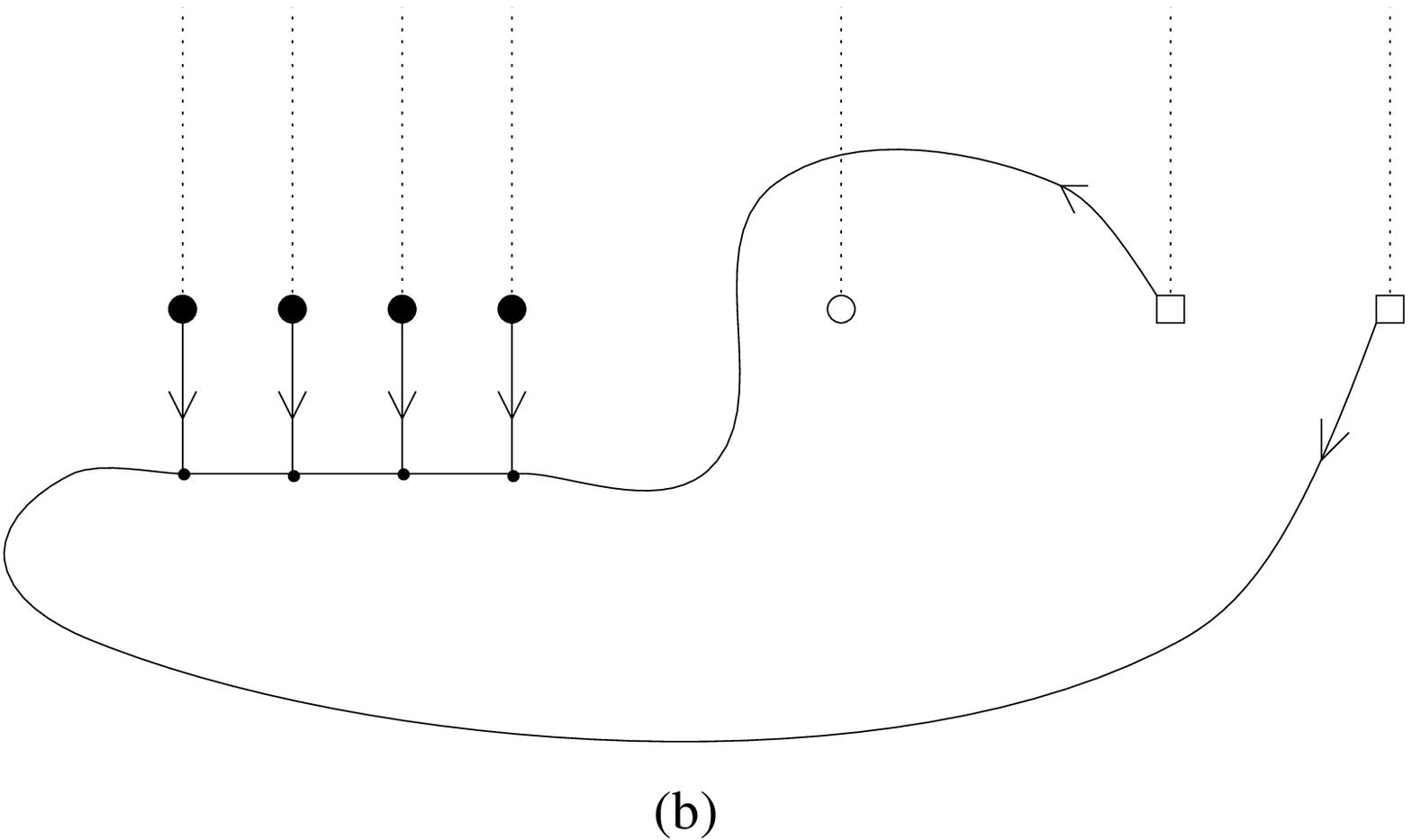}}
\vspace{0.3cm}

\epsfysize=4.0cm
\centerline{\epsffile{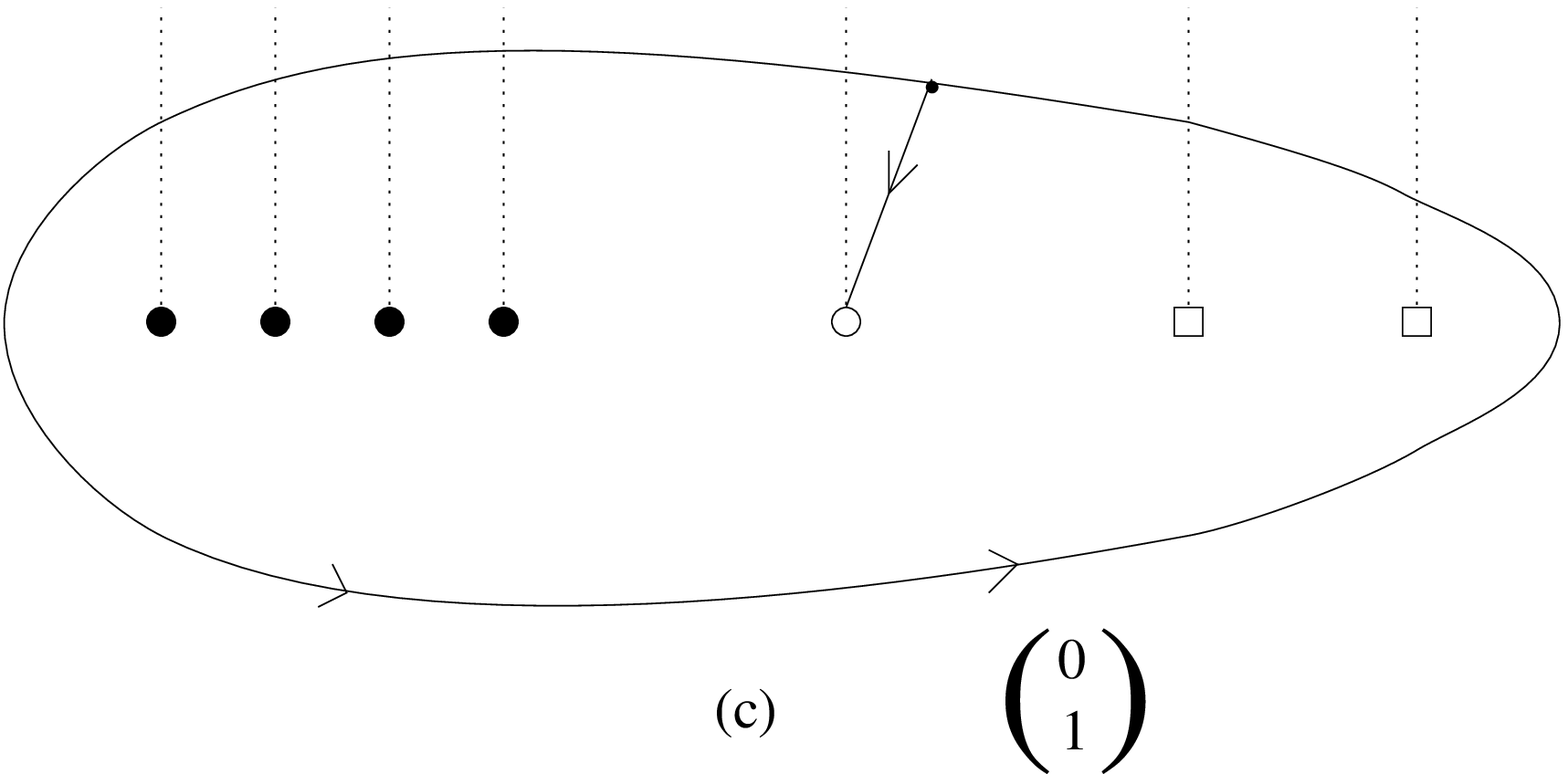}}
\caption{Different versions of the indirect $CC$ geodesic which transforms
as $(\square^4_a, \cdot )$ of $su(n_A) \times su(n_C)$.}
\end{figure}

The charge with respect to the $A$ branes can be made manifest by pushing
the open string across the $A$ branes, thereby creating four open
string prongs (Fig.~4(b)). The four prongs have the same direction and
represent the antisymmetric part of the fourth power of the
fundamental as can be seen by composing these configurations with direct
$AA$ strings.
\smallskip

In part (c) of figure 4 we illustrate a somewhat surprising presentation
of the indirect $CC$ states, where a single open string prong attaches
the $B$ brane to a closed string loop by means of a three string junction. 
The fact that this is an allowed diagram follows, algebraically, from the
fact that $({0\atop 1})$ is left invariant by $A^4B^2C^2$.\footnote{The 
$B^2$ factor arises because by the time the $({0\atop 1})$ string is
about to reach the junction, it has become a $({2\atop -1})$ string, and the
effect of the $B$ prong is the same as that of crossing a $B$ cut.}
In figure~5 we show a couple of steps that demonstrate that this 
representation can be related by our moves to Fig.~4 (a):
as a first step we move the string in Fig.~4 (c) across the two $C$ branes
thus creating two prongs. The point where the left prong joins the closed
sting can be slided all the way until it stands to the left of the $A$
cuts.  At this stage the upper part of the closed string can be pushed down
across the $B$ brane, thereby eliminating the $B$ prong, and across the $A$
branes without creating prongs. After these steps the string looks like
that in Fig.~5(b). The loop can be collapsed and we recover the familiar
indirect $CC$ geodesic of Fig.~4(a).

\begin{figure}[htb]
\epsfysize=4.1cm
\epsffile{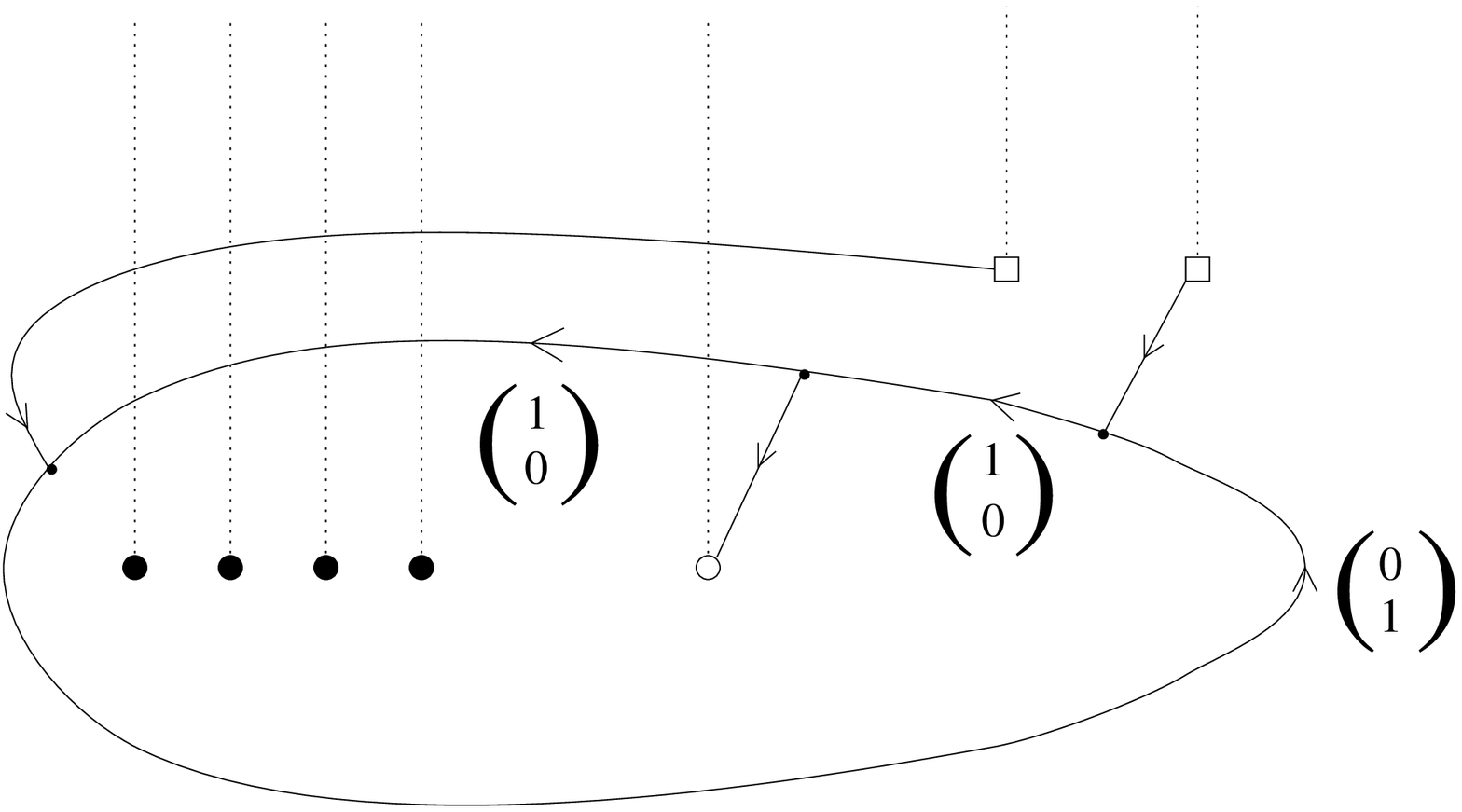}
\vspace{-4cm}

\epsfysize=4.5cm
\hfill{\epsffile{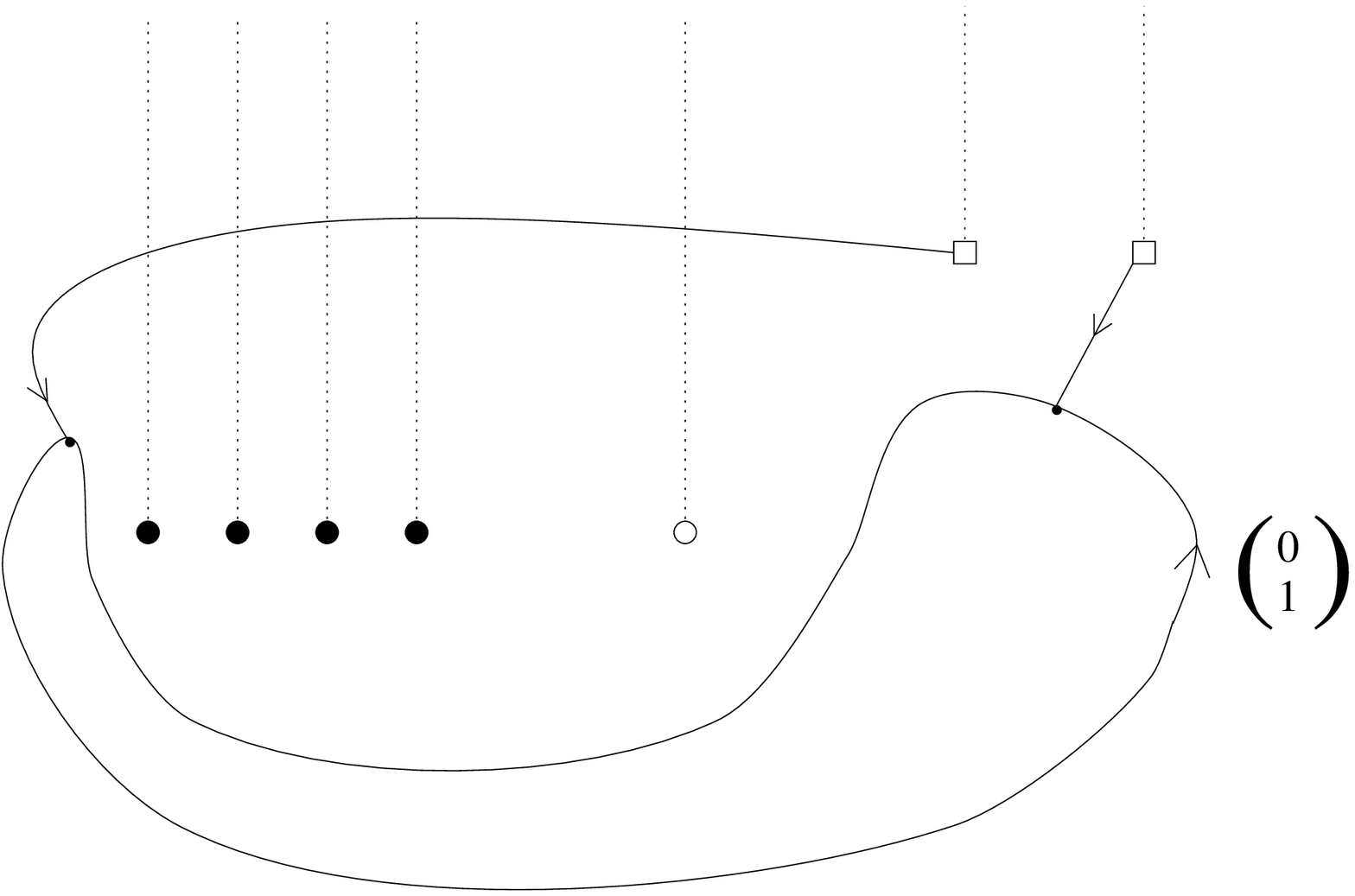}}
\caption{ The left figure is a modified version of Fig.~4 (c). A couple of
moves relates it to the right figure which is recognized as equivalent to
Fig.~4 (a).} 
\end{figure}

\subsection{Construction of  $E_6$}

In the previous section we have discussed the relevant multi-pronged open
strings that give rise to the representations $(10,2)_{-3}$ and
$(\overline 5, 1)_{-6}$ that together with their complex conjugates are
necessary to enlarge the manifest subalgebra 
$su(5)\times su(2)\times u(1)$ to $E_6$. In 
this section we verify, using the various representatives we have
discussed, that the multi-pronged open strings can be combined by the usual
rules of joining open strings to give representatives in the class of the
expected product.  We will not discuss all possible products we may form,
but rather illustrate how the rules work by means of two non-trivial
examples. We should also mention that whilst we are able to generate all
necessary generators, it is less clear why these are all generators, and
why other (seemingly possible) diagrams are not relevant. This is
presumably a difficult problem whose solution would require a better
understanding of the BPS condition in this context. On the other hand, this
problem is not really new: for example in the above description of $so(8)$,
the $AA$ geodesic that winds twice around the $BC$ singularity does not
correspond to a gauge vector, and further ``selection rules'' are
necessary.
\medskip

In the first example we discuss the product $(10,2)_{-3}\times
(\overline{10},2)_{+3}$, {\it i.e.} the multiplication of $(\square^2_a,
\square)$ with its conjugate; this is done in Fig.~6.  As a first step we
release the strings from the common $A$ brane and collapse it partially by
crossing a $C$ brane and creating a prong. This gives Fig.~6 (b).  We then
move the string across the second $C$ brane, create a second prong, and
obtain Fig.~6 (c). The junctions joining each of the $C$ prongs to the
other string can be collapsed, and a direct open string joining the two $C$
branes can be separated out.  The final result is a direct $AA$ string and
a direct $CC$ string, and represents the fact that the product contains
{\it both} the $(\hbox{Adj} , \cdot )$ and the $(\cdot, \hbox{Adj})$ of
$su(n_A)\times su(n_C)$.

\begin{figure}[htb]
\epsfysize=5cm
\epsffile{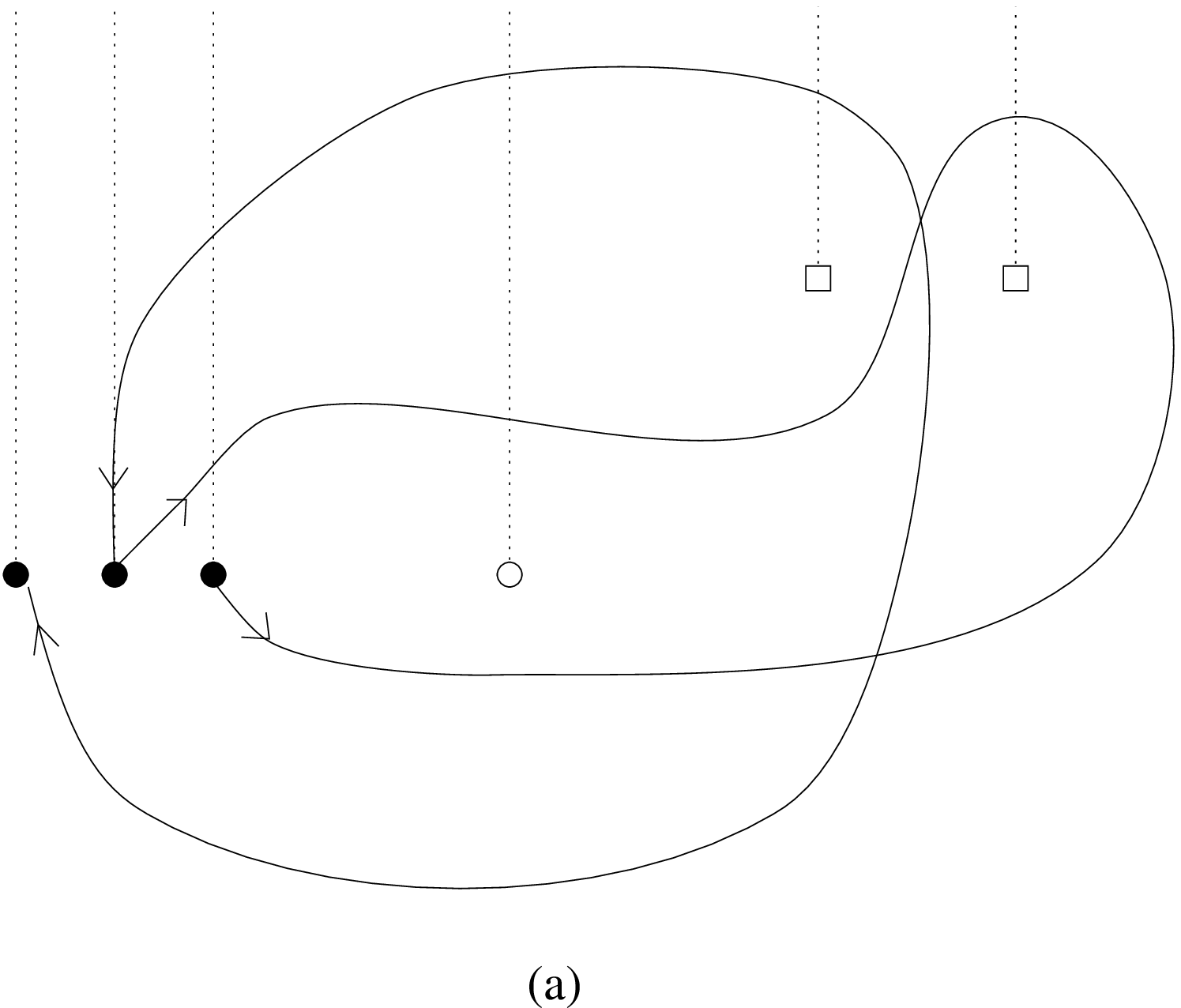}
\vspace*{-5cm}

\epsfysize=5cm
\hfill{\epsffile{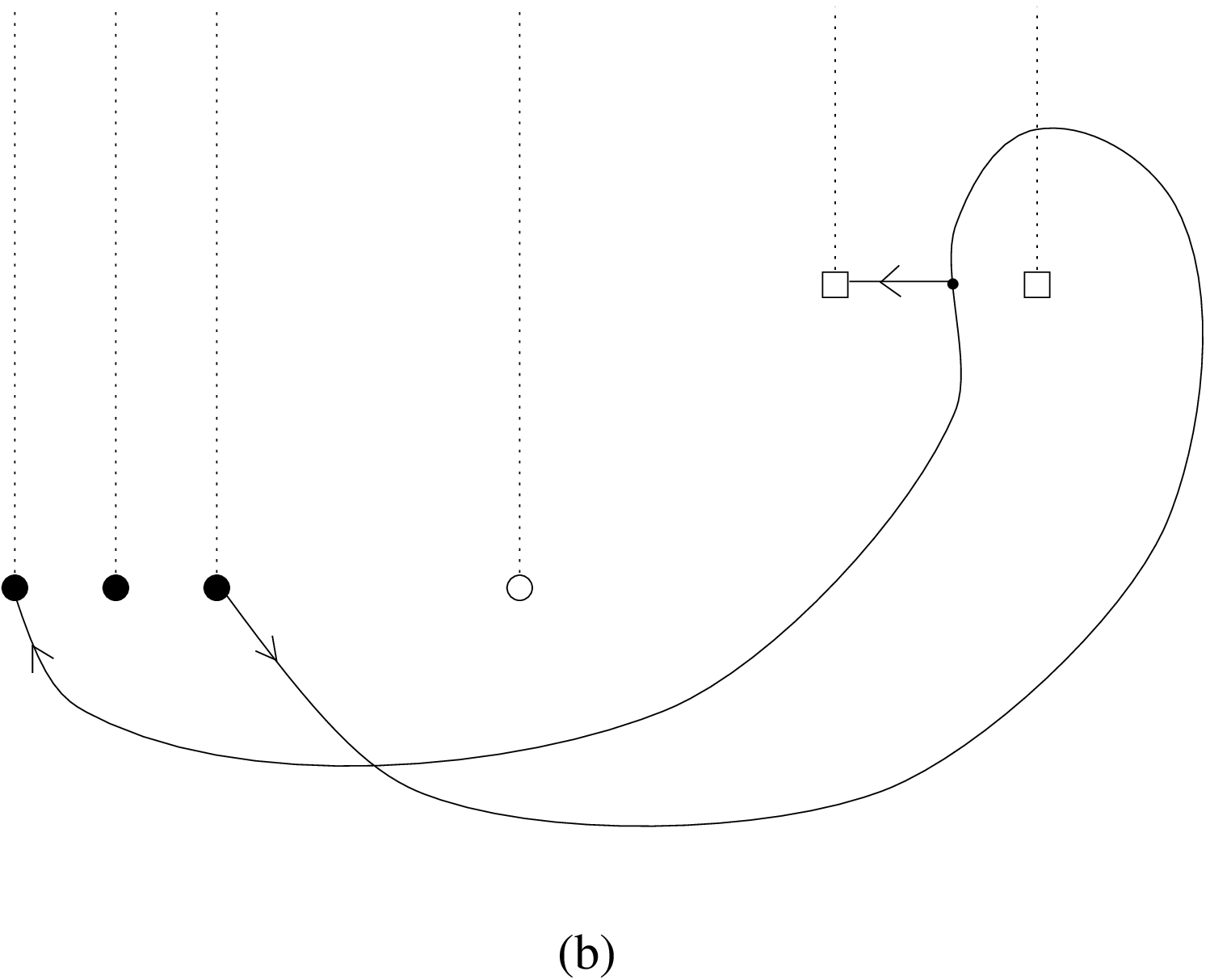}}
\vspace{0.3cm}

\epsfysize=5cm
\epsffile{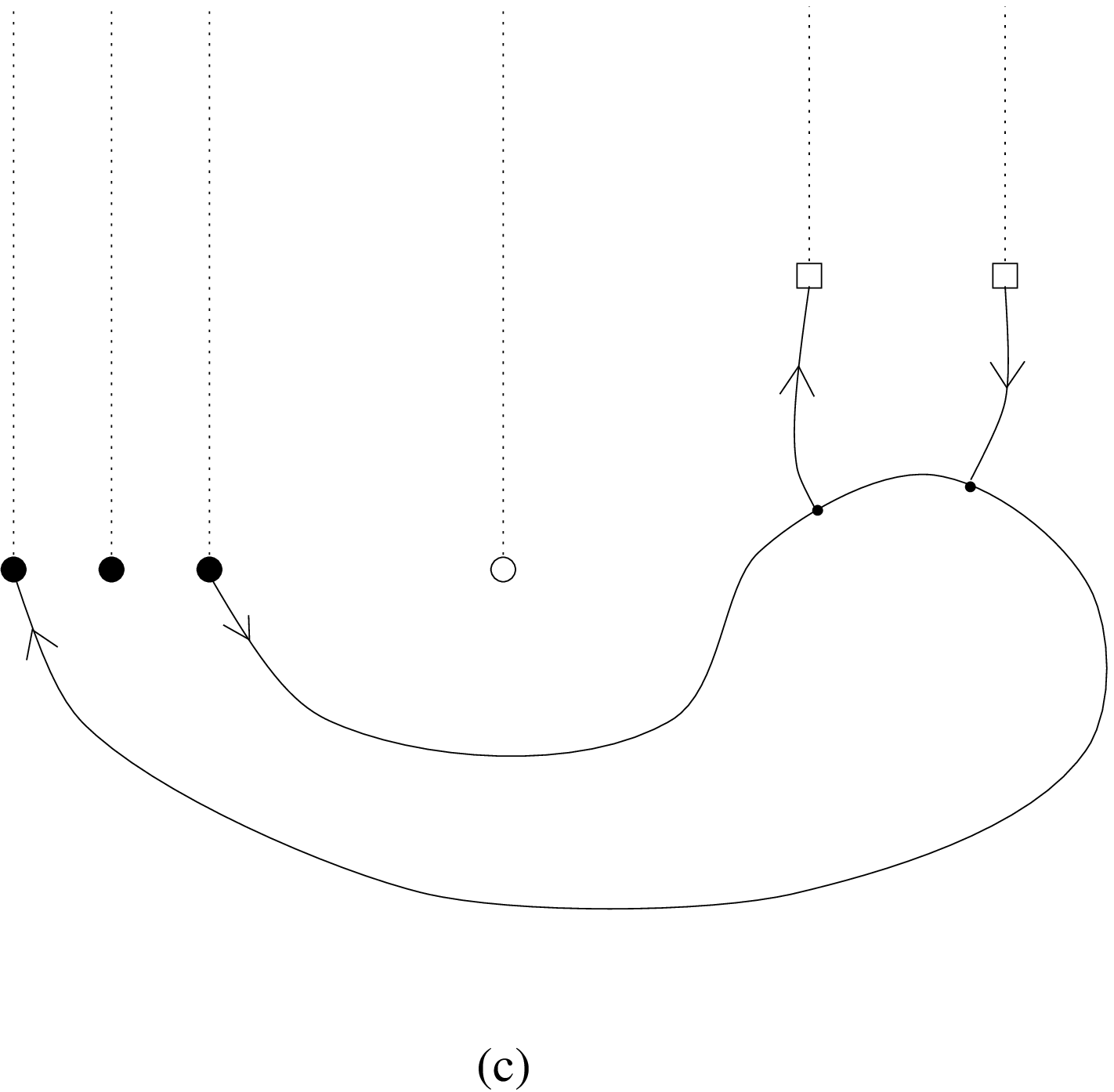}
\vspace*{-4.4cm}

\epsfysize=4cm
\hfill{\epsffile{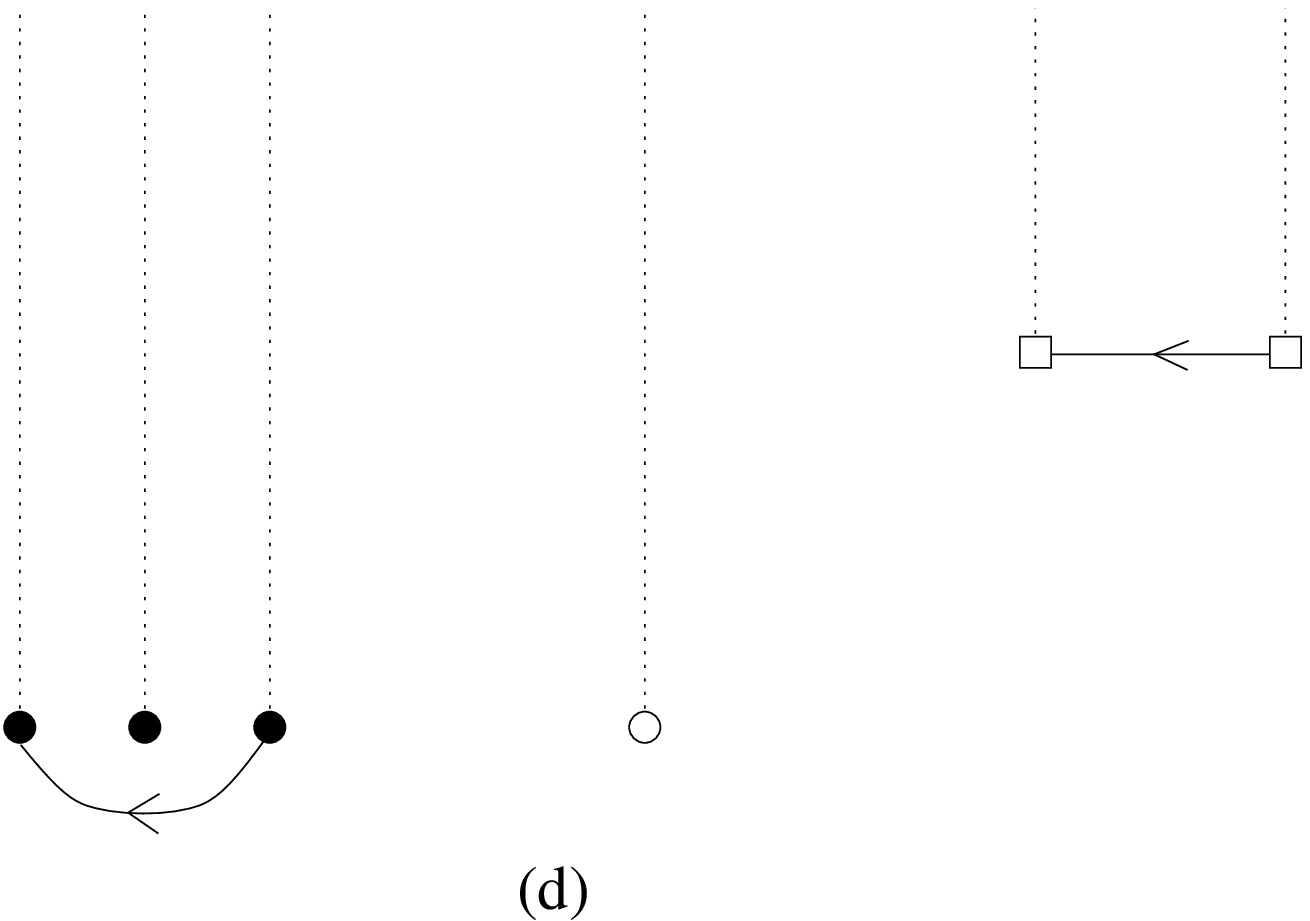}}
\vspace*{0.8cm}

\caption{ Multiplication of $(\square^2_a, \square )$ times its
conjugate. The final result gives two representations, the 
$(\hbox{Adj},\cdot)$ and the $(\cdot, \hbox{Adj})$ of 
$su(n_A) \times su(n_C)$.} 
\end{figure}

In the second example we discuss the product $(10,2)_{-3}\times
({10},2)_{-3}$, {\it i.e.} the product of $(\square^2_a, \square )$ with
itself. It is convenient to choose representatives carefully to facilitate
the computation, and we use representatives (d) and (e) of  Fig.~3, as is
shown in Fig.~7. The prongs ending on the $B$ brane can be combined and
released, and the result is immediately recognized as the 
$(\square^4_a,\cdot )$ string in the representation  shown in Fig.~4 (b). 
This, of course, is as it should be. 

\begin{figure}[htb]
\epsfysize=4cm
\centerline{\epsffile{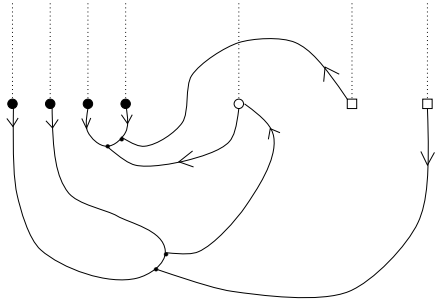}}
\caption{Multiplication of $(\square^2_a,\square )$ of 
$su(n_A) \times su(n_C)$ with itself. The result gives the
$(\square^4_a, \cdot)$ of $(su_A, su_C)$.}
\end{figure}

\subsection{Construction of  $E_7$}

For the case of $E_7$, as we have seen in the table, we have one
representation that has no analogue in $E_6$, the $(1,2)_{-3}$
representation which corresponds to $(\square^6_a, \square)$ of
$su(6)\times su(2)$. Since there are only five $A$ branes in the $E_6$
case, this representation is not present.  
\smallskip

In Fig.~8 we show explicitly how to find open string representatives for
these generators. We construct them by multiplying $(\square^2_a,\square )$ 
with $(\square^4_a, \cdot)$. For $(\square^2_a,\square )$ we choose
the representative (c) of Fig.~3, where the bottom part of the string has
been pushed upwards to cross the $B$ brane and create a prong; for 
$(\square^4_a, \cdot)$ we use Fig.~4 (a). These two representatives join
suitably at one of the $C$ 
branes, where the string can be released. Its junction on 
the upper string can then be slided to the left crossing a $C$ cut, a $B$ cut and
four $A$ cuts. At this stage a move allows one to enclose the fifth $A$ brane
(counting from the right), and we find the result of Fig.~8 (b). The open 
string prong at $B$ can then be traded via a move with a configuration
where the remaining $A$ brane is now enclosed, giving us the result shown 
in Fig.~8 (c). This is a reasonable presentation as an intricate but
conventional open string.  Its transformation properties under $su(2)$ are
manifest as the string ends on a $C$ brane, and the nature of the
presentation is not changed by composition with a direct $CC$ open
string.\footnote{A different representative was proposed in
Ref.\cite{johansen}. That representative does not appear to transform
properly under $su(2)$ as composition with a direct $CC$ string gives a
string of different nature.}
\smallskip

An interesting and useful representative can be found starting from
Fig.~8 (b), and this time pushing the string across the leftmost $A$ brane
giving us the configuration indicated in Fig.~8 (d). The $B$ prong can
then be removed by lowering the string across the $B$ brane, and after
sliding the junction one finds the result indicated in (e). This is a one
pronged object made of a closed loop and an open string joining it to a $C$
brane. It is the analogue of the presentation of Fig.~4 (c) for the $CC$
states. Its transformation properties under $su(n_A)$ and $su(n_C)$ are
manifest.

\begin{figure}[htb]
\epsfysize=3.4cm
\epsffile{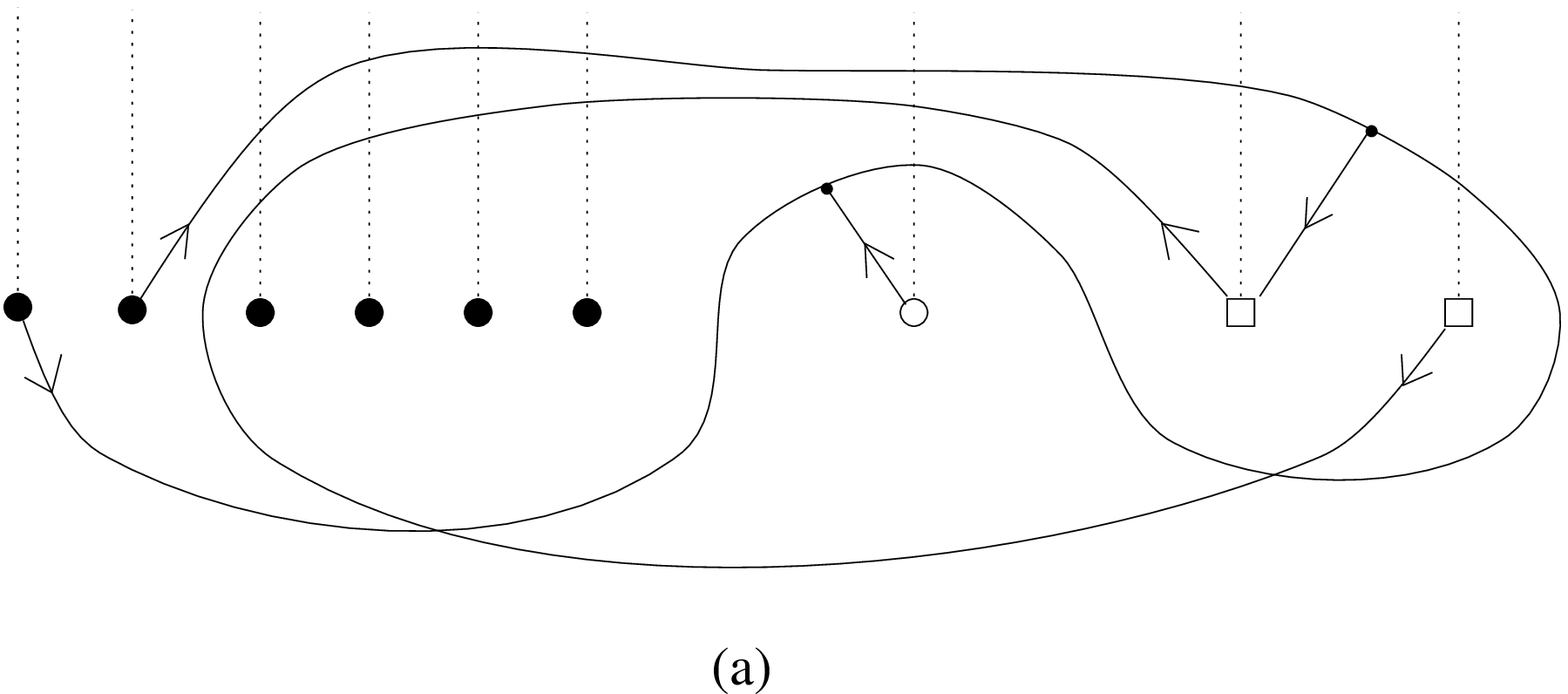}

\epsfysize=3.4cm
\hfill{\epsffile{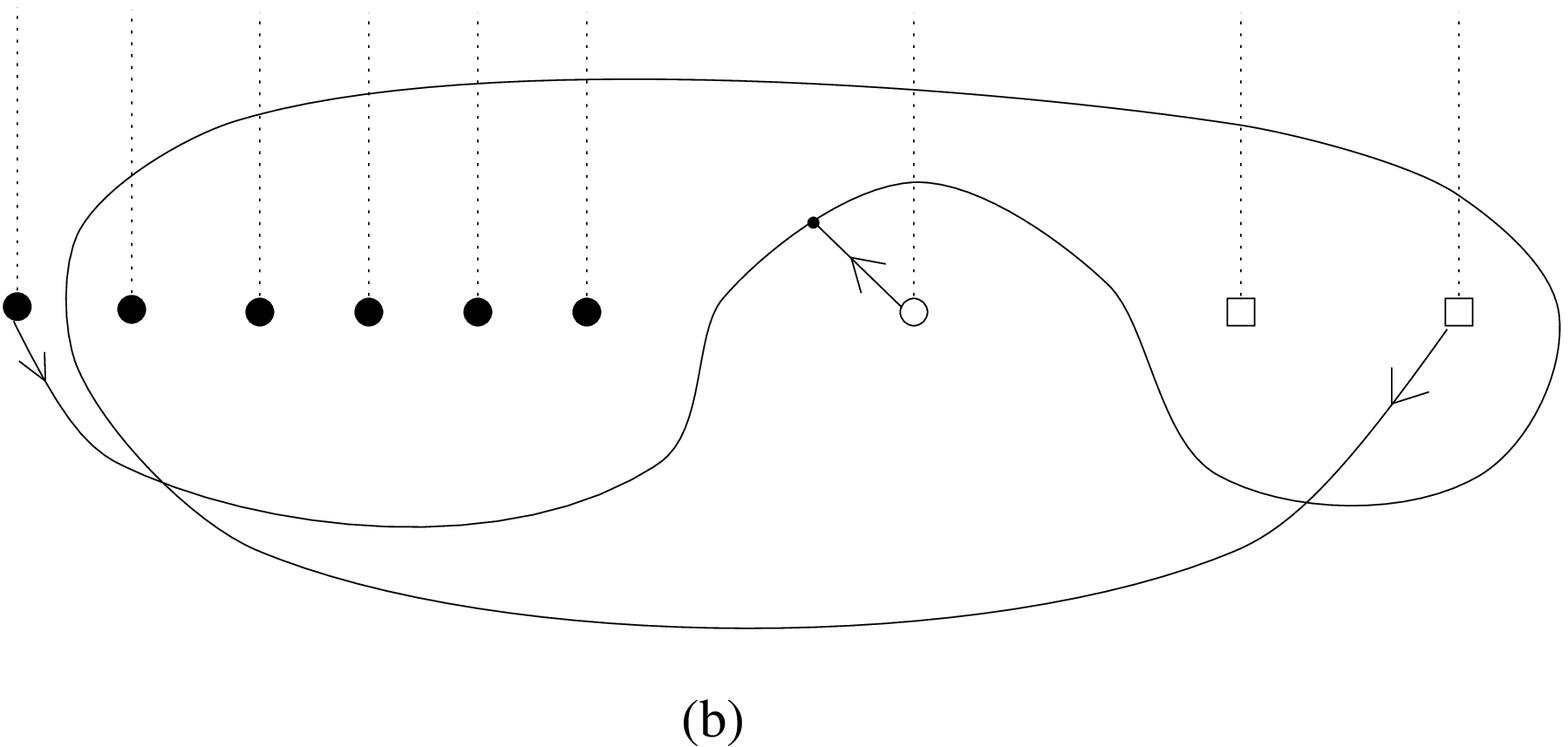}}

\epsfysize=3.4cm
\epsffile{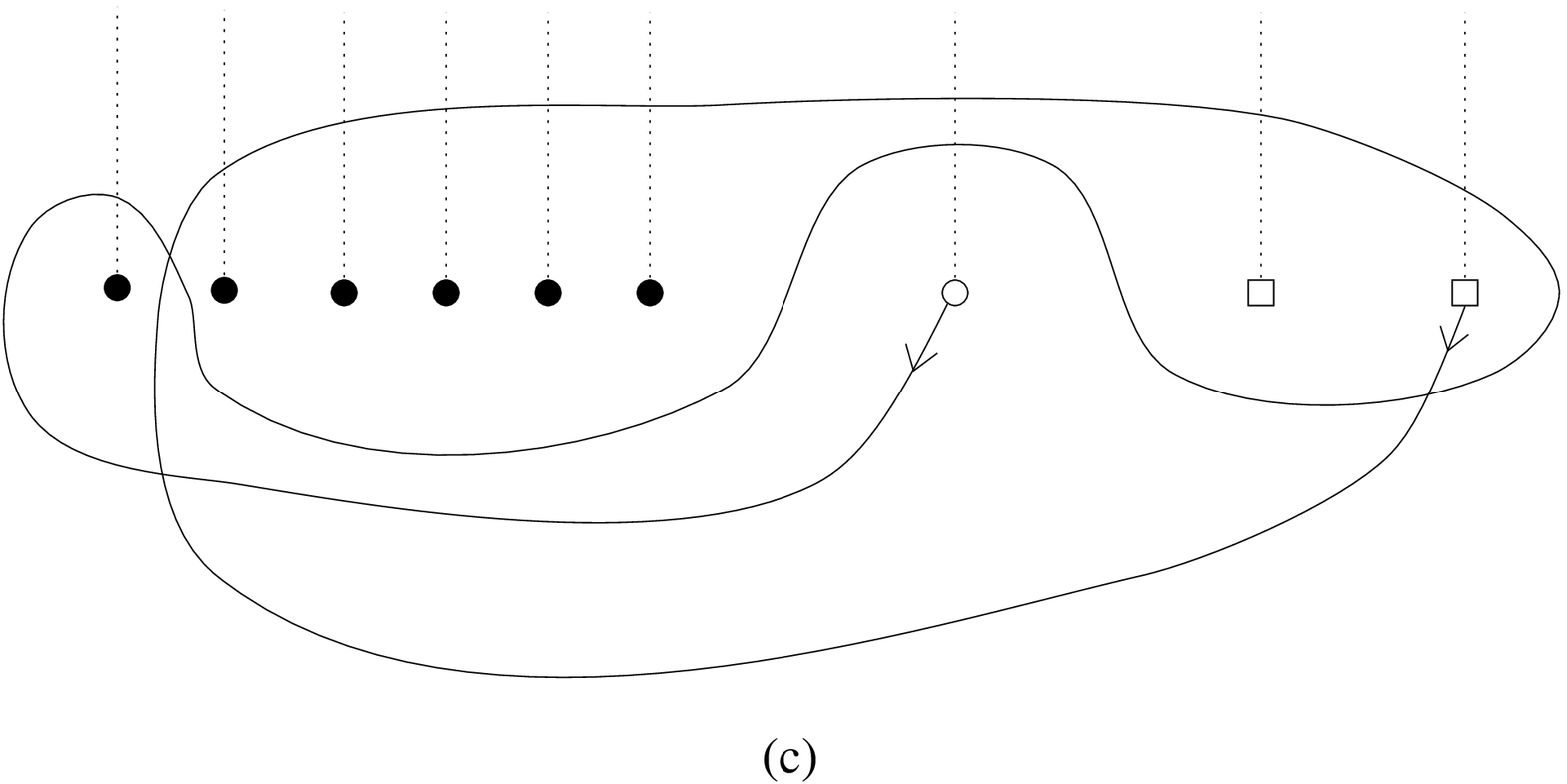}
\vspace{0.8cm}

\epsfysize=3.4cm
\hfill{\epsffile{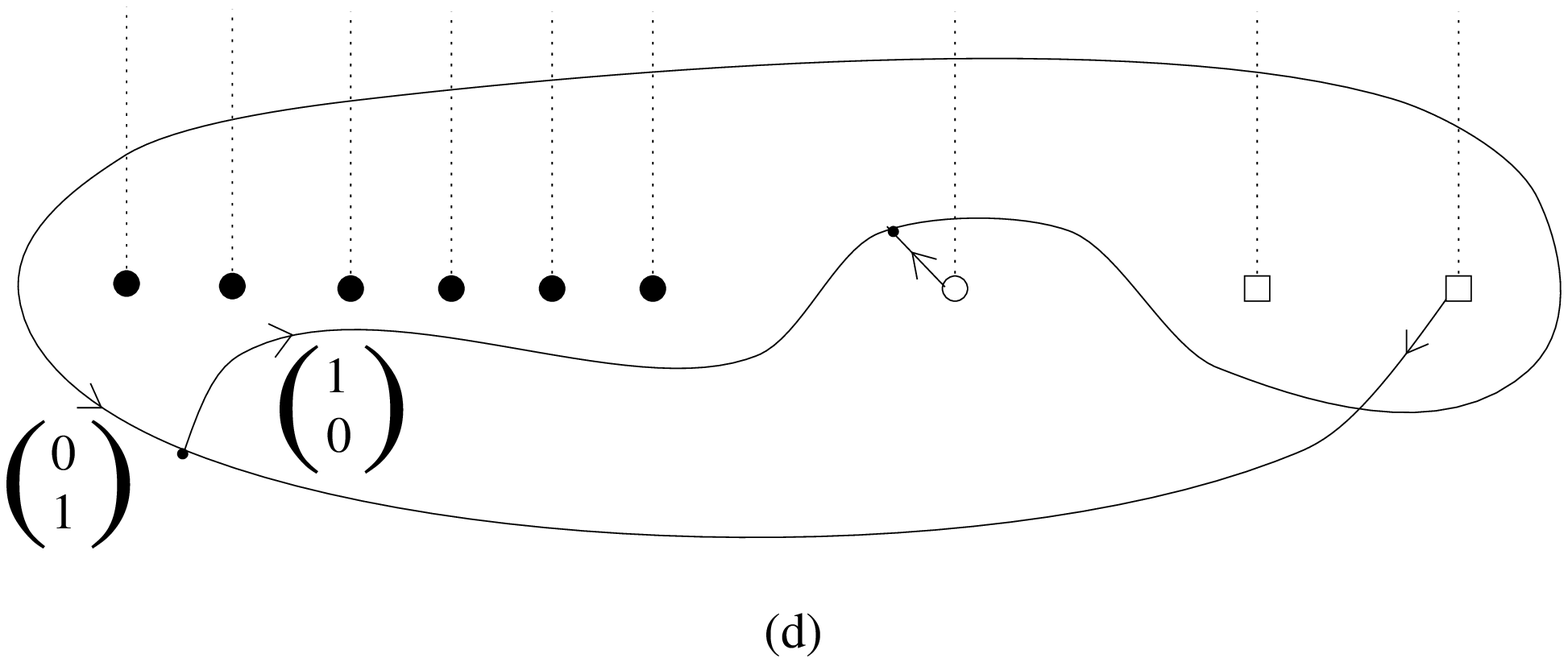}}
\vspace{1.4cm}

\epsfysize=3.4cm
\centerline{\epsffile{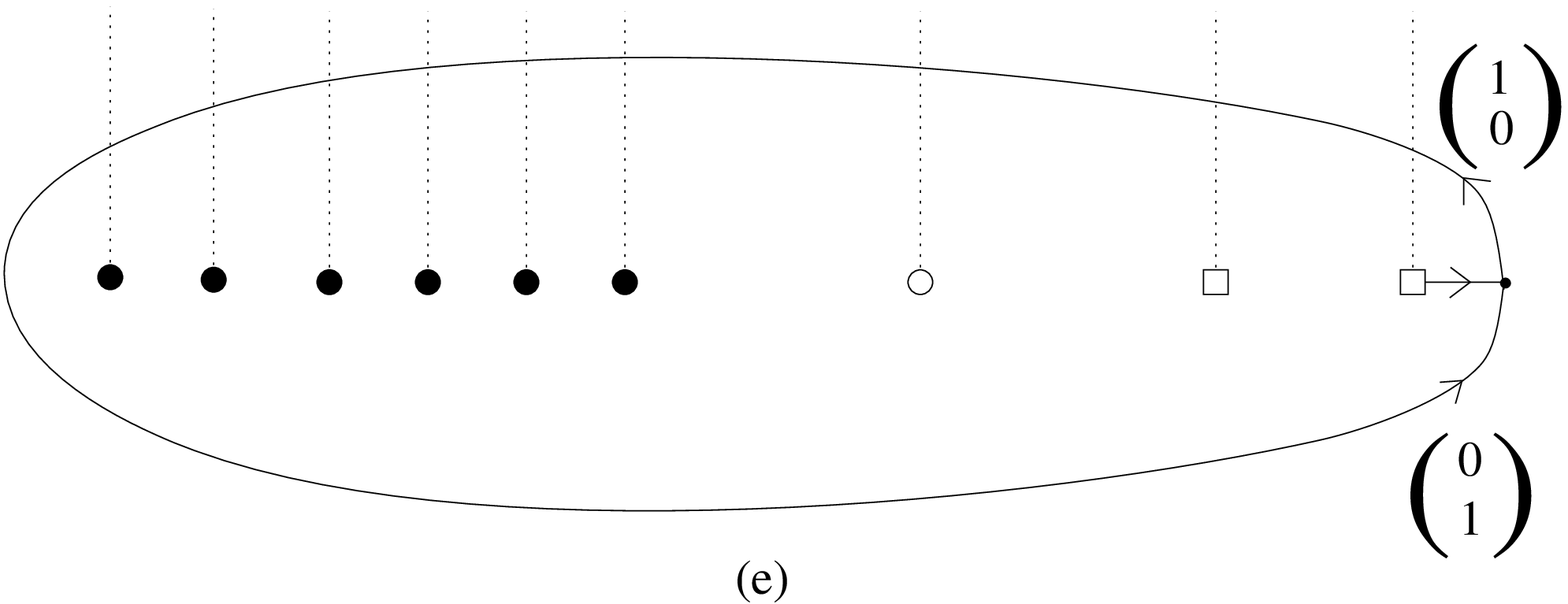}}
\caption{Multiplication of $(\square^2_a,\square )$ times 
$(\square^4_a, \square)$ of $su(n_A) \times su(n_C)$.}
\end{figure}

\subsection{Construction of  $E_8$}

For the case of $E_8$, as we have seen in the table, there exists
one representation that has no analogue for $E_7$, the
$(7,1)_{8}$ which transforms as $(\square , \cdot )$.
In Fig.~9 we find an explicit open string representative for this
representation by multiplying the $(\square^2_a, \square )$ and 
$(\square^6_a, \square)$ of $su(7) \times su(2)$. For the 
$(\square^2_a, \square )$ states we use the presentation of Fig.~3 (c), and
for $(\square^6_a, \square)$ the presentation of Fig.~8 (e). These two
representatives join suitably at one of the $C$ branes, where the string
can be released. Moreover we can move the left part of the string across
the leftmost $A$ brane, creating a junction and finding the representation
shown in Fig.~9 (b). We can slide this junction all the way to the right, as
in Fig.~9 (c), and continue to slide it until it hits the other junction,
at which time we can collapse the resulting loop and obtain the
presentation shown in Fig.~9 (d). Once more, we have found a one pronged
object made of a closed loop and an open string, joining it to an $A$
brane. Its transformation properties under $su(n_A)$ and $su(n_C)$ are
manifest.

\begin{figure}[htb]
\epsfysize=3.5cm
\centerline{\epsffile{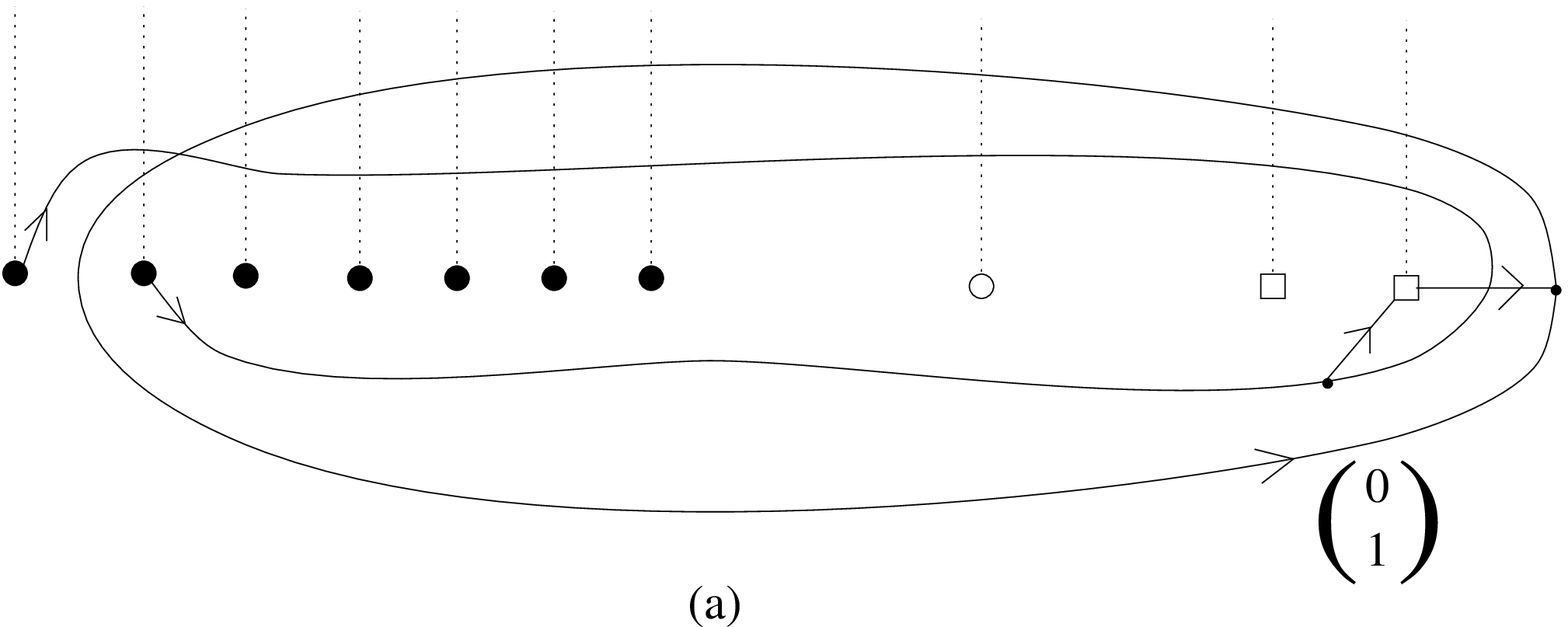}}
\vspace*{1.0cm}

\epsfysize=4.0cm
\centerline{\epsffile{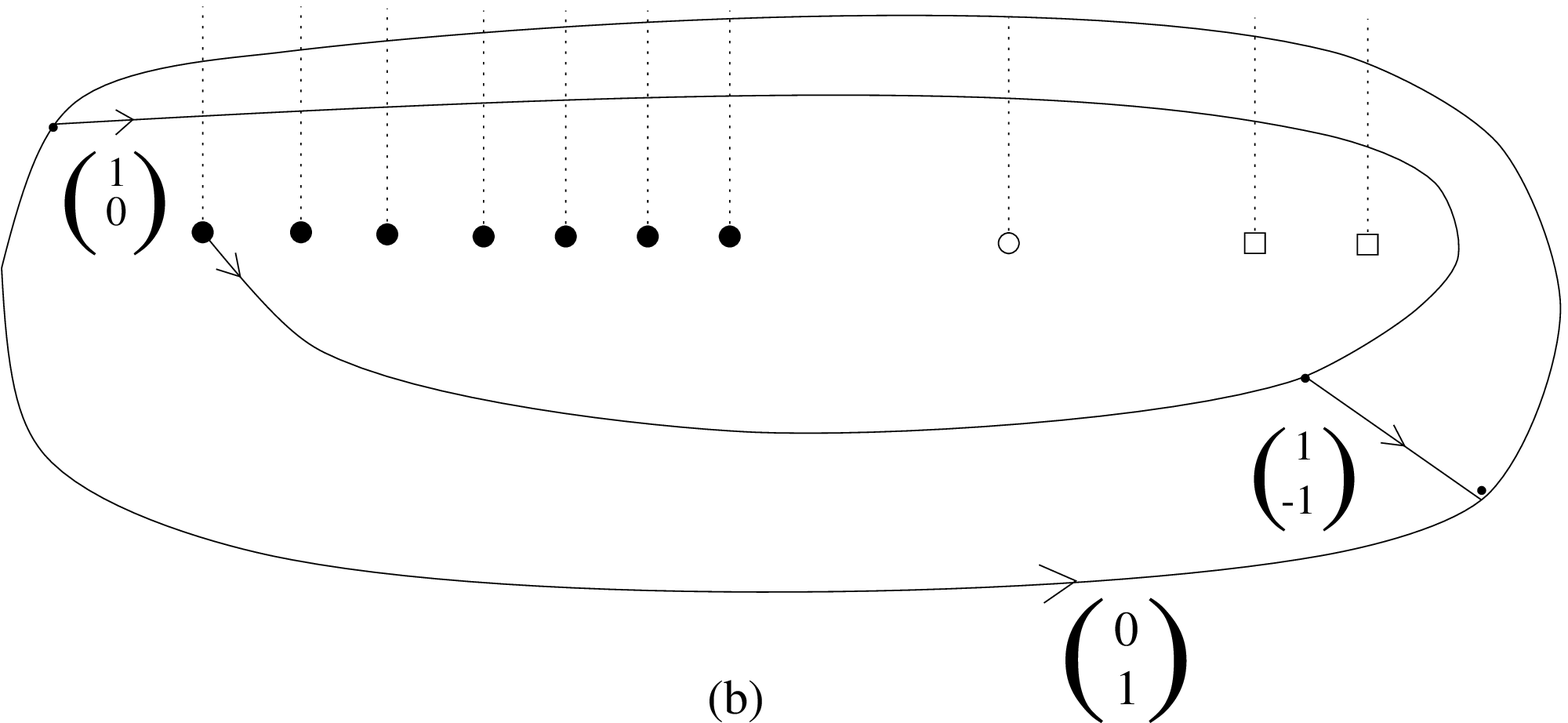}}
\vspace*{1.0cm}

\epsfysize=4.2cm
\centerline{\epsffile{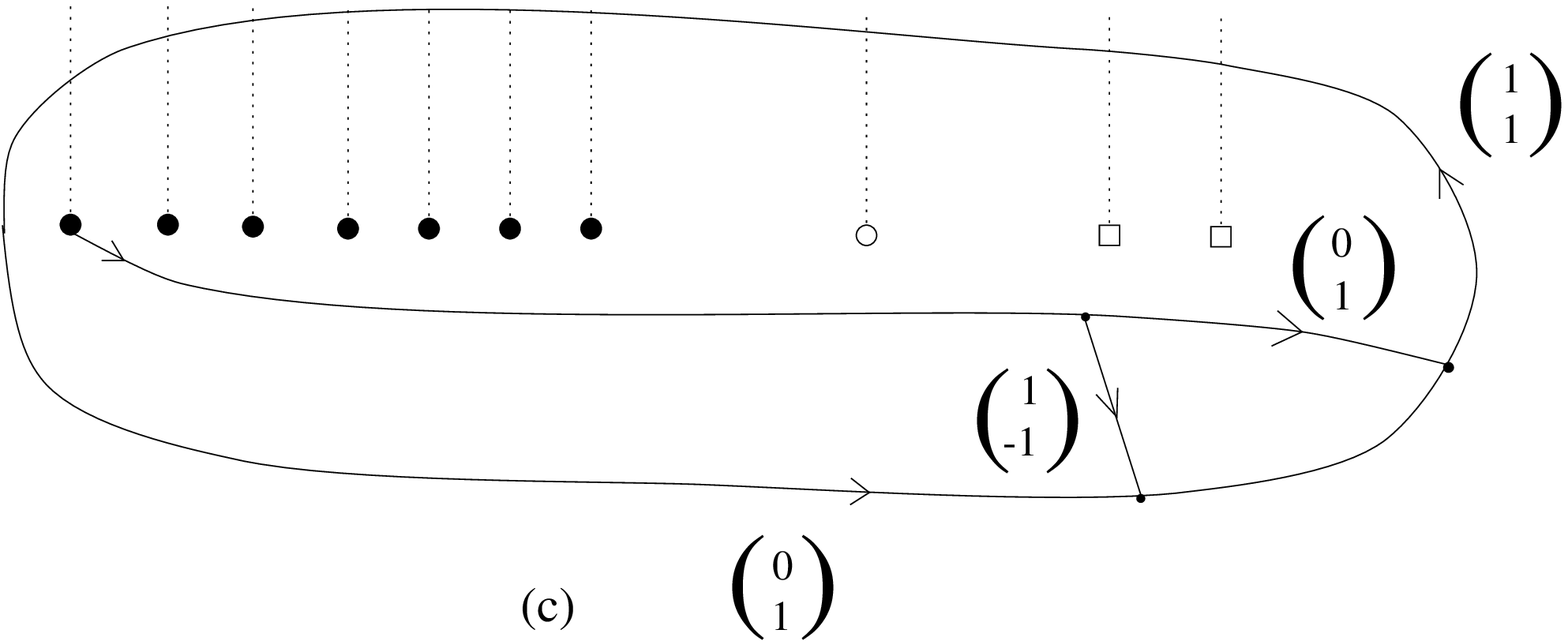}}
\vspace*{1.5cm}

\epsfysize=3.8cm
\centerline{\epsffile{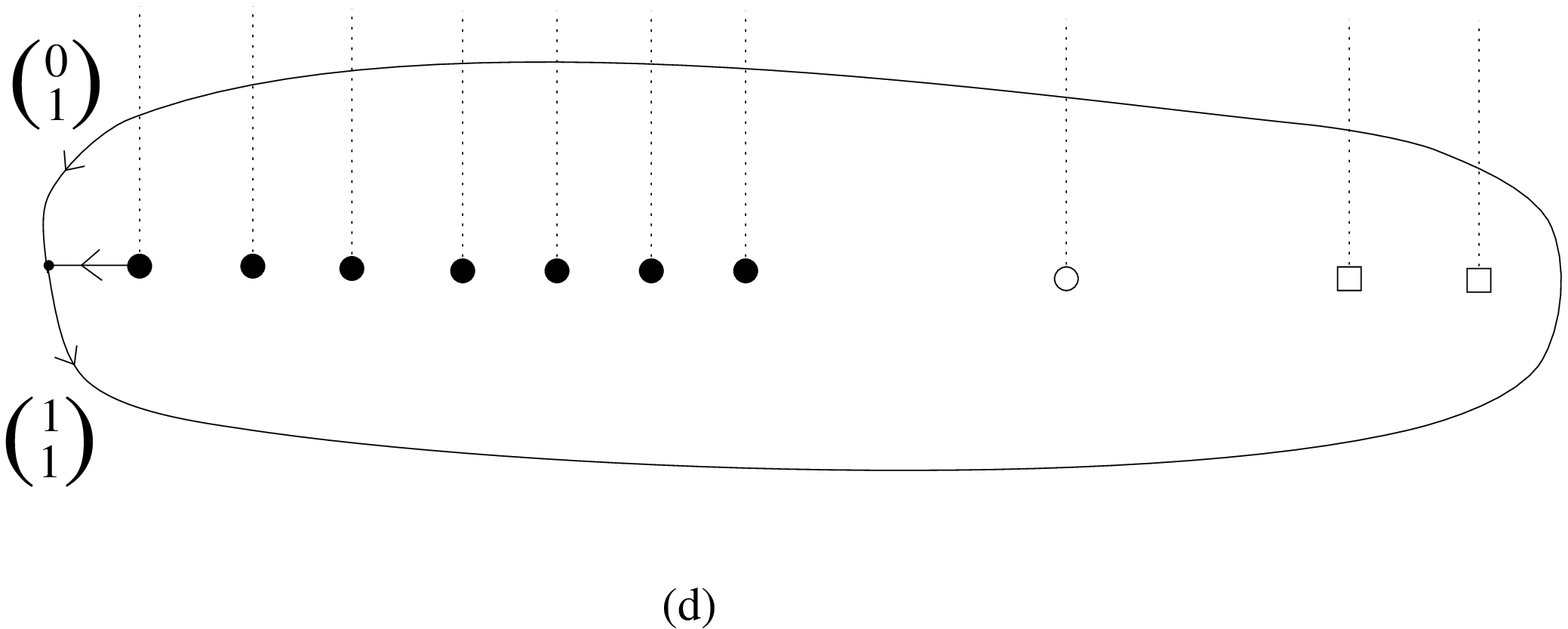}}
\vspace*{0.5cm}

\caption{Multiplication of the representations $(\square^2_a, \square )$ 
and $(\square^6_a, \square)$ of $su(n_A) \times su(n_C)$. For $E_8$ where
$su_A = su(7)$ the answer contains the $(\square, \cdot)$.}
\end{figure}

\section{Conclusions and open questions}

We have shown in this paper that an open string interpretation for the
gauge vectors of the exceptional Lie algebras $E_6, E_7$ and $E_8$ can be
given if multi-pronged open string are included as well as conventional
open strings.  The charge assignments and the transformation properties of
the various states are manifest, and strings can be combined correctly by
the conventional operation of joining open strings, or more precisely, open
string prongs. We have also shown how some of the junction diagrams can be
related to conventional open string diagrams using a modification of the
Hanany-Witten effect.
\medskip

We have seen that our method is powerful enough to generate very simple
open string representations for the states whose conventional open string
representations were not known.  It is particularly striking that all
states that are necessary to go beyond classical Lie algebras can be
represented as closed string loops with an open string attaching it to a
particular brane. $E_6$ requires such a loop attached to the $B$ brane,
$E_7$ to the $B$ brane, and the $C$ branes, and $E_8$ to the $A$, $B$ and
$C$ branes.
\medskip

We have demonstrated that these configurations account for the necessary
gauge vectors; however, as in previous analysis of classical groups, it is
less clear under which conditions a particular open string geodesic or
multi-pronged open string corresponds to a BPS state, and thus to a gauge
vector.  
\medskip

We believe that the open string junctions are more than useful mathematical
devices to account pictorially for the intricacies of the representation
theory of exceptional Lie algebras. In fact, our analysis suggests that the
multi-pronged open strings are the correct realization of BPS states in some
regions of the moduli space of 7-brane positions. As we move in moduli
space the representative of the BPS state can change, and we expect it to
do so according to the crossing rules we have explained.  If this physical
interpretation is confirmed by further exploration, a tantalizing
non-perturbative picture of open string theory including all kinds of open
string junctions emerges.

\section*{Acknowledgments}

\noindent We wish to thank M. Bershadsky who brought to our attention the
earlier work on exceptional algebras and D-branes.  We are grateful to
A. Johansen for explanations on his work, and to C. Vafa for suggestions
that were very helpful in the analysis of geodesics.
\smallskip

\noindent M.R.G. is supported by a NATO-Fellowship and in part by NSF grant
PHY-92-18167.  B.Z. is supported in part by D.O.E.  contract
DE-FC02-94ER40818, and a fellowship of the John Simon Guggenheim Memorial
Foundation.

\pagebreak

\end{document}